\documentclass[prb,superscriptaddress,amsmath,amssymb,twocolumn]{revtex4-2}

\usepackage[usenames,dvipsnames]{xcolor}
\usepackage{amsmath,amssymb}
\usepackage{graphicx}
\usepackage{hyperref}
\usepackage{braket}
\usepackage[normalem]{ulem}
\usepackage{enumerate}
\usepackage{MnSymbol}

\newcommand{\be}{\begin{equation}}
\newcommand{\ee}{\end{equation}}

\newcommand{\dd}{{\rm d}}

\newcommand{\sm}{{s_{\text{m}}}}
\newcommand{\q}{{\rm q}}

\newcommand{\dr}{{\text{dr}}}
\newcommand{\ddr}{{\textbf{dr}}}
\newcommand{\eff}{{\text{eff}}}

\newcommand{\gI}{{I}}

\begin{document}

\newcommand{\titleinfo}{
The sine-Gordon model from coupled condensates: \\ a 
Generalized Hydrodynamics viewpoint}

\title{\titleinfo}

\author{Alvise Bastianello}
\affiliation{Technical University of Munich, TUM School of Natural Sciences, Physics Department, 85748 Garching, Germany}
\affiliation{Munich Center for Quantum Science and Technology (MCQST), Schellingstr. 4, 80799 M{\"u}nchen, Germany}

\begin{abstract}
The sine-Gordon model captures the low-energy effective dynamics of a wealth of one-dimensional quantum systems, stimulating the experimental efforts in building a versatile quantum simulator of this field theory and fueling the parallel development of new theoretical toolkits able to capture far-from-equilibrium settings.
In this work, we analyze the realization of sine-Gordon from the interference pattern of two one-dimensional quasicondensates: we argue the emergent field theory is well described by its classical limit and develop its large-scale description based on Generalized Hydrodynamics. We show how, despite sine-Gordon being an integrable field theory, trap-induced inhomogeneities cause instabilities of excitations and provide exact analytical results to capture this effect.
\end{abstract}

\maketitle

\section{Introduction}
Over the last decades, much emphasis has been put on understanding quantum many-body physics out of equilibrium, driven both by the per-s\'e interest in fundamental research, and for developing quantum technologies.
In this quest, low-dimensional systems cover a pivotal role \cite{Bloch2008,Bloch2012}.
In parallel with experimental progress, equally groundbreaking theoretical achievements have been made, both on the numerical and analytical side. 
Of course, the most successful approach surges from combining these strategies, whenever possible, but such fortuitous instances are rare, and each theoretical approach comes with its own limitations.

On the numerical side, tensor-network algorithms \cite{Verstraete2008} are extremely versatile in describing the evolution of low-entangled states, but they are constrained in time and excitations' energy. In contrast, exact diagonalization is severely limited in the system's size. Finally, semiclassical methods such as the Truncated Wigner Approximation \cite{Polkovnikov2010} may miss important quantum effects.

The challenge of analytically solving an interacting many-body system can be tackled only in a handful of special cases, but, when possible, it keeps up to the promise of outmatching ab-initio numerical methods: integrability \cite{Guan2022} is the ideal playground for fulfilling this program.
Experimental realizations of integrable models are always imperfect: even when the local interactions of the target model are eingeneered correctly, large-scale inhomogeneities caused by the trapping potential holding the atoms are unavoidable and clash against canonical methods of Bethe Ansatz \cite{Smirnov1992}. Hence, the capability of several experiments of studying far-from-equilibrium protocols remained beyond the reach of both numerical and analytical means for a long time.

The tables turned with the advent of Generalized Hydrodynamics (GHD)\cite{Bertini2016,Doyon2016,specialissueGHD}: this hydrodynamic theory builds on the extensive number of conservation laws featured by integrable systems and exactly derscribes their large-scale dynamics, capturing the effect of inhomogeneities through generalized force fields \cite{Doyon2017,Bastianello2019int,Durnin2021}. This leap forward brought experiments within theoretical reach and GHD has been readily proven capable of describing trap quenches in one-dimensional Bose Gases \cite{Schemmer2019,Malvania2020,Moller2021,Cataldini2022,Bouchoule2022} which, to date, stands out as the only instance where an actual experiment has been compared with GHD predictions.

However, the arena of integrable models realized in the lab is much broader: multicomponent generalizations of the Bose Gas and their fermionic counterparts are natural candidates \cite{Guan2013,Senaratne2022}, and experimental investigations of Heisenberg magnets with tunable anisotropy have started \cite{Hild2014,Jepsen2020,Jepsen2021,Scheie2021,Wei2022}. Applications of GHD to these experiments can be foreseen in the forthcoming future \cite{Bulchandani2021,Cecile2022,Scopa2021,Scopa2022}.

The sine-Gordon (SG) model is the integrable field theory par-excellence with the broadest applicability in condensed matter \cite{Zvyagin2004,Umegaki2009,Essler1998,Essler1999,Lomdahl1985,Davidson1985, Roy2019,Roy2021}.
A versatile experimental realization \cite{Gritsev2007,Gritsev2007a} is engineered in Vienna by J. Schmiedmayer's group \cite{Schweigler2017,Zache2020,Pigneur2018}: the tunability of this setting opens the door to investigate many aspects of this field theory, both in and out of equilibrium, and this experiment became a reference point for a broad community.
It is appealing to investigate this experimental setup through the lenses of GHD, which may prove extremely useful in going beyond the current theoretical understanding, mainly based either on non-interacting limits \cite{Yuri2020a,Foini2015,Foini2017,Ruggiero2021,Yuri2018}, self-consistent gaussian approximations \cite{Yuri2019}, semiclassical numerical analysis \cite{DallaTorre2013,Kasper2020} or Truncated Conformal Space Approach (TCSA) \cite{Feverati1998,Horvath2019,Horvath2022,szaszschagrin2023,Kukuljan2018,Kukuljan2020} that generalizes exact diagonalization methods to the modes of the field theory.
While TCSA is able to capture quantum effects far from equilibrium, it is challenging to include inhomogeneities and the large scales typical of the experimental setup.

The development of a hydrodynamic treatment of sine-Gordon is largely untapped, with so far only studies of transport in homogeneous backgrounds \cite{Bertini2019,Koch2023}: the underlying inhomogeneous atomic cloud has crucial effects on the sine-Gordon dynamics, leading to force fields and, most importantly, quasiparticle instabilities \cite{Bastianello2019,Koch2021,Koch2022}.

In this work, we derive the sought hydrodynamic equations within the semiclassical regime of the field theory, completing the study of the classical sine-Gordon initiated in Ref. \cite{Koch2023}.
The motivations are multifaceted: first, we argue that the parameter range explored by the current experiment is very close to the semiclassical limit, avoiding the need for a quantum treatment. Second, the semiclassical regime can be benchmarked against Monte Carlo simulations, backing up analytical calculations. Lastly, these results will serve as a stepping stone for future developments in the even more challenging quantum realm \cite{Nagy2023}.

This work is so organized. In Section \ref{sec_the_model}, we provide an overview of sine-Gordon and its exact thermodynamics, emphasizing when the field theory is well-approximated by its classical limit.
In Section \ref{sec_experiment} we discuss the quasicondensates experimental realization and the applicability of the semiclassical approximation.
Section \ref{sec_ghd} gathers the central result of this work, namely the Generalized Hydrodynamics of the classical sine-Gordon: we discuss the hydrodynamic equations and provide numerical evidence of their correctness.
Our conclusions are collected in Section \ref{sec_conclusions}, where we also discuss the open challenges and future directions. A few technical appendixes follow.

\begin{figure}[t!]
\begin{center}
\includegraphics[width=1 \columnwidth]{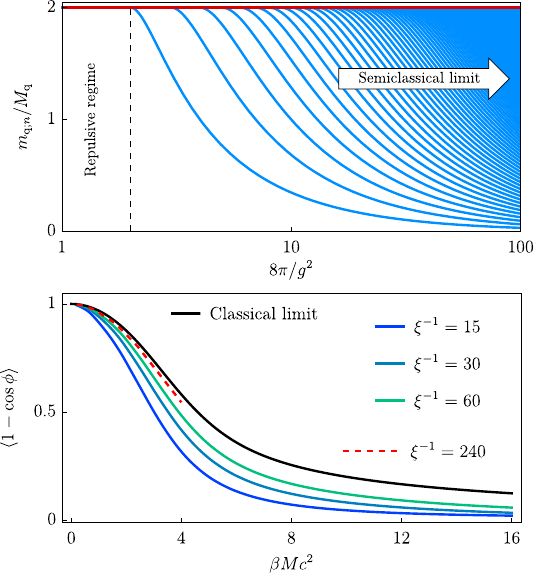}
\end{center}
\caption{\textbf{The scaling to the semiclassical limit.} 
Top: We plot the mass law of breathers, renormalized to the soliton mass, for different values of the interaction $g$. As $g$ is decreased, more and more breathers appear in the spectrum and the mass-quantization is blurred into a continuum, collapsing to the semiclassical limit. See Section \ref{sec_1_a} for further discussion.
Bottom: We consider $\langle 1-\cos\phi\rangle$ as a function of the adimensional quantity $\beta M c^2$, by comparing the classical result (black line) with quantum curves for different values of the interaction. For the sake of simplicity, we focus on the so-called reflectionless points $\xi^{-1}\in \mathbb{N}$ of sine-Gordon. The semiclassical limit is achieved for $\xi^{-1}\to\infty$: we show the extrapolation of quantum curves to $\xi^{-1}=240$ (dashed red line) due to its experimental relevance, see Section \ref{sec_experiment}.
For a discussion of the quantum value of $\langle 1-\cos\phi\rangle$ and of the extrapolation, see Appendix \ref{app_quantum_TBA}.}\label{Fig_1}
\end{figure}

\section{The sine Gordon model}
\label{sec_the_model}

The sine-Gordon model is a relativistic field theory governed by the following Hamiltonian
\be\label{eq_HSG}
H=\int\dd x\, c\left\{\frac{g^2}{2}\Pi^2+\frac{1}{2 g^2}(\partial_x\phi)^2+\frac{m^2 c^2}{g^2}(1-\cos \phi)\right\}\, .
\ee
Throughout this work, we use units such that the Planck constant is one $\hbar=1$.
Above, $\Pi(t,x)$ is the field conjugated to the phase $\phi(t,x)$, the free parameters of the theory are the light velocity $c$, the bare mass scale $m$ and the interaction $g$.

The SG model is integrable both in the classical \cite{novikov1984,Faddeev1987} and quantum \cite{Zamolodchikov1979} formulation: in the latter, classical fields are promoted to bosonic operators $[\phi(x),\Pi(x')]=\delta(x-x')$. . In this section, we discuss the rudiments of both these regimes, and the connection between the two. For the sake of notation, we add a label ``q" to quantum objects that differ from their classical counterparts.

We begin with the classical theory \cite{novikov1984}: the fundamental excitations are topological in nature and interpolate between the degenerated vacua of the potential $\phi=2\pi \mathbb{Z}$. Kinks are excitations with a positive phase winding of $2\pi$ going from left to right in the spatial direction. Antikinks are obtained by spatial inversion and thus diminish the phase of a $2\pi$ unit. Kinks and antikinks are relativistic particles with the same dispersion law, i.e. they have energy $\epsilon_K(\theta)=Mc^2\cosh\theta$ and momentum $p_K(\theta)=Mc \sinh\theta$, where $\theta$ is the rapidity and $M=8 m/g^2$ the soliton mass.
These excitations are also present in the quantum case \cite{Zamolodchikov1979}: here,  quantum effects renormalize the kinks' mass leading to $M_q\propto m^{1+\xi}$
\cite{Zamolodchikov1995,Lukyanov1997}, where $\xi^{-1}=8\pi/g^2-1$, see also Appendix \ref{app_quantum_TBA} for the full expression.

Due to interactions, kink-antikink pairs form stable boundstates called ``breathers", whose dispersion law is exactly known: in the quantum realm, breathers are labeled by an integer quantum number $n=\{1,2,...\}$, and the number of species is crucially determined by the interaction $n< \xi^{-1}$. Their mass-scale follows the simple law $m_{n;\q}=2M_\q\sin(\frac{\pi}{2}n\xi)$ \cite{Zamolodchikov1995}. In particular, for $8\pi/g^2<1$, the quantum sine-Gordon is unstable under renormalization group \cite{Zamolodchikov1995}, for $1<8\pi/g^2<2$ sine-Gordon is within the repulsive phase, where only kinks and antikinks are present in the spectrum, but breathers are absent. For $8\pi/g^2\ge 2$, breathers of increasingly many species are possible as $g$ is diminished, see also Fig. \ref{Fig_1} (top). 

Within the classical realm, breathers are no longer quantized and they always belong to the spectrum, labeled by a continuum spectral parameter $s \in [0,\sm]$ and mass law $m_s=M\sin\left(\frac{\pi}{2}\frac{s}{\sm}\right)$ \cite{Faddeev1987}, where we defined $\sm=8\pi/g^2$. In the literature, it is often conventional to use a different parametrization for the spectral parameter $\sigma\in[0,1]$ by defining $\sigma=s/\sm$, but we found the first choice to be more convenient for us.
These quasiparticles exhaust the list of the low-energy excitations of the sine-Gordon model: its finite temperature thermodynamics can be then built using Thermodynamic Bethe Ansatz (TBA) \cite{takahashi2005thermodynamics} in the quantum case and its analog formulation in the classical realm \cite{Currie1980,El2021}. 

Thanks to integrability, finite-energy states in the systems, both quantum or classical, can be understood in terms of asymptotic multiparticle states, built from a collection of low-energy excitations with elastic scattering. 
Upon scattering, these excitations have an effective length that depends on the rapidities of the particles, due to the Wigner time delay or scattering shift \cite{Doyon2018}. In the classical SG, two breathers with relative rapidity $\Delta\theta$ experience a scattering shift $\varphi_{s,s'}(\Delta\theta)$ \label{novikov1984}
\begin{multline}\label{eq_varphi}
\varphi_{s,s'}(\Delta \theta)=\frac{16}{g^2}\log\left(\frac{\cosh \Delta\theta-\cos((s+s')\tfrac{\pi}{2\sm})}{\cosh\Delta\theta+\cos((s+s')\tfrac{\pi}{2\sm})}\right)+\\
\frac{16}{g^2}\log\left(\frac{\cosh \Delta\theta+\cos((s-s')\tfrac{\pi}{2\sm})}{\cosh \Delta\theta-\cos((s-s')\tfrac{\pi}{2\sm}}\right)\, .
\end{multline}

The kink(antikink) scattering shift with a breather is obtained as a limit of the above expression $\varphi_s(\Delta\theta)=\frac{1}{2}\lim_{s'\to \sm}\varphi_{s,s'}(\Delta\theta)$, and similarly the scattering shift between two topological excitations $\varphi(\Delta\theta)$ (symmetric over kink-antikink exchange) is recoved by a further limit $\varphi(\Delta\theta)=\frac{1}{2}\lim_{s\to \sm}\varphi_{s}(\Delta\theta)$.
Scattering shifts in the quantum model are also exactly known \cite{Zamolodchikov1979} and have more complicated expressions (see also Appendix \ref{app_quantum_TBA}), but their exact form is not needed here. As an important difference, in the quantum SG the scattering of kink-antikink can be both transmissive and reflective, whereas it is always transmissive in the classical case.

The excitations' dispersion laws and the scattering shifts are at the basis of the exact thermodynamics, that we now overview.

\subsection{The classical thermodynamics}
\label{sec_cl_thermo}

Despite the thermodynamics of the classical sine-Gordon model has been intensively studied for years \cite{Timonen1986,Currie1980,Takayama1985,Chen1986,Maki1985,Chung1989,Kazuo1986,Theodorakopoulos1984b,Chung1990}, its correct formulation became available only very recently \cite{Koch2023}.
Since we find convenient a slight change of the notation compared to the original reference, we quickly recap the main formulae.

Thermal states, and more generally Generalized Gibbs Ensembles \cite{Calabrese2016}, of the classical sine-Gordon are identified by the so-called root densities, one for the breathers $\rho_s(\theta)$, for the kinks $\rho_{K}(\theta)$ and antikinks $\rho_{\bar{K}}(\theta)$. The root density can be interpreted as the phase space density of the excitations, similar to the soliton-gas picture \cite{El2021}. The non-trivial scattering among these particles renormalizes the phase-space, hence one also introduces the total root densities $\rho^t_s(\theta)$, $\rho^t_{K}(\theta)$ and $\rho^t_{\bar{K}}(\theta)$ and finally the filling functions defined as the ratio of the two $\vartheta_\gI(\theta)=\rho_\gI(\theta)/\rho_\gI^t(\theta)$, for each particle species. From now on, we use the generic label ``$\gI$" whenever we give general statements holding for breathers, kinks and antikinks.

It is also useful to introduce the dressing operation. 
To this end, it is convenient to use a vector-matrix notation: any triplet of functions $\{\tau_s(\theta),\tau_K(\theta),\tau_{\bar{K}}(\theta)\}$ is organized into a vector $\tau$  in the space of rapidities and internal degrees of freedom $s$, $K$, and $\bar{K}$. We introduce a kernel $\varphi$ acting on this space and a convolution $\star$, in such a way $\varphi\star\tau=\{(\varphi\star\tau)_s(\theta),(\varphi\star\tau)_K(\theta),(\varphi\star\tau)_K(\theta)\}$, where the components read
\begin{multline}
(\varphi\star\tau)_s(\theta) =\int\frac{\dd\theta'}{2\pi}\Big[\varphi_{s}(\theta-\theta')(\tau_K(\theta')+\tau_{\bar{K}}(\theta')]+\\
\int\frac{\dd\theta'}{2\pi}\int_0^{\sm} \dd s'  \varphi_{s, s'}(\theta-\theta')\tau_{s'}(\theta')\Big]\, ,
\end{multline}
\begin{multline}
(\varphi\star\tau)_K(\theta) =\int\frac{\dd\theta'}{2\pi}\Big[\varphi(\theta-\theta')(\tau_K(\theta')+\tau_{\bar{K}}(\theta')]+\\
\int\frac{\dd\theta'}{2\pi}\int_0^{\sm} \dd s'  \varphi_{s'}(\theta-\theta')\tau_{s'}(\theta')\Big]
\end{multline}
and a similar definition holds for the antikinks.

With this notation, we define the dressing operation $\tau\to \tau^\dr$ as the solution of the linear equation
\be
\tau^\dr=\tau-\varphi\star [\vartheta\tau^\dr]\, ,
\ee
where the product of vectors between square brackets is meant to be taken on each component $[\vartheta\tau^\dr]_\gI(\theta)=\vartheta_\gI(\theta)\tau^\dr_\gI(\theta)$.
The root densities and total root densities are not independent quantities, but they are connected through the dressing of the rapidity-derivative of the momentum as $\rho^t_\gI(\theta)=\frac{1}{2\pi}(\partial_\theta p_\gI)^\dr$.

On thermal states, integral equations for the filling functions have been derived in Ref. \cite{Koch2023}. In the notation of this work, we can write them as
\begin{multline}\label{eq_TBA}
-\log(s^2\vartheta_s(\theta))+2s/\sm=\beta \epsilon_s(\theta)+\\\int\frac{\dd\theta'}{2\pi}\varphi_s(\theta-\theta')(\vartheta_K(\theta')+\vartheta_{\bar{K}}(\theta'))+\\
\int\frac{\dd\theta'}{2\pi}\int_0^{\sm}\dd \,s'\varphi_{s,s'}(\theta-\theta')\frac{(s')^2\vartheta_{s'}(\theta')-1}{(s')^2}\, .
\end{multline}
and $\vartheta_K^2(\theta)=\vartheta_{\bar{K}}^2(\theta)=\lim_{s\to\sm}\vartheta_s(\theta)$. 

We notice that the filling function obtained as a solution of Eq. \eqref{eq_TBA} has a singular behavior for small strings $\vartheta_s\sim 1/s^2$, therefore on practical implementations it is more convenient to reparametrize the equations in terms of the non-singular part, see Appendix \ref{app_GHDdiscretization}.

Once the filling functions and root densities have been obtained, observables can be computed. Of particular interest for us is the cosine of the phase, which reads \cite{Koch2023}

\begin{multline}\label{eq_HF}
2 \frac{m^2 c^2}{g^2}\langle 1-\cos\phi\rangle=\int \frac{\dd\theta}{2\pi}  \int_0^\sm \dd s \, (c^{-1}\epsilon_s\epsilon_s^\dr-c p_sp^{\dr}_s )\vartheta_s+\\
\int \frac{\dd\theta}{2\pi}  \Big\{\, (c^{-1}\epsilon_K\epsilon_K^\dr-c p_Kp^{\dr}_K )\vartheta_K+(c^{-1}\epsilon_{\bar{K}}\epsilon_{\bar{K}}^\dr-c p_{\bar{K}}p^{\dr}_{\bar{K}} )\vartheta_{\bar{K}}\Big\}\, .
\end{multline}

This is a convenient measure of interactions and temperature, and it is experimentally accessible \cite{Schweigler2017}, as we will see in Section \ref{sec_experiment}.

\subsection{From quantum to classical}
\label{sec_1_a}

Semiclassical limits of quantum field theories \cite{Blakie2008} are generally achieved as a combination of weak interactions and large occupation numbers, eventually resulting in large temperatures. While this scaling does not rely on integrability, the latter allows for a careful study of the limit by comparing exact results at finite temperature in the quantum and classical cases \cite{DeLuca2016,Bastianello2018}, quantifying the validity of the approximation.

One can already grasp the basics from the mass spectrum depicted in Fig. \ref{Fig_1}: the classical limit is achieved for small interactions $g$, when the discrete quantum spectrum is well-approximated as a continuum. Furthermore, the energy of two consecutive breathers should be indistinguishable on the energy scale determined by the inverse temperature $\beta$, i.e. one asks $\beta [\epsilon_{n+1;\q}(\theta)-\epsilon_{n;\q}(\theta)]\ll 1$, resulting in the two conditions
\be\label{eq_semiclassic}
\frac{g^2}{8\pi}\ll1\hspace{1pc}\text{and}\hspace{1pc}\beta mc^2\cosh\theta\ll 1\, ,
\ee
where we also used that the quantum soliton mass becomes the classical one in the limit. In this regime, it holds the correspondence $s\leftrightarrow n$ between the quantum breathers' index $n$ and the classical spectral parameter $s$. With this rescaling, and at fixed rapidities, the quantum scattering shift becomes the classical one upon taking $g$ small \cite{Koch2023}.

At large rapidities, Eq. \eqref{eq_semiclassic} cannot be fulfilled, making quantum effects important in the UV limit or equivalently at short distances $\ell_\text{UV}\lesssim \beta  c$.
However, realistic experiments have a finite resolution and effectively put a cap on the minimum length that can be probed, possibly preventing access to the deep UV limit. It remains to be seen if the UV part of the spectrum may affect finite-momentum excitations when considering thermodynamics, but this is not the case: the scattering shift between an excitation with small rapidity and another in the UV part of the spectrum decays as $\sim 1/\cosh\Delta\theta\sim 1/\cosh\theta\lesssim \beta m c^2$, hence the UV part of the spectrum decouples from the rest.
In practice, when is the semiclassical approximation valid?

To answer this question, in Fig. \ref{Fig_1} (bottom) we compare $\langle 1-\cos\phi\rangle$ \eqref{eq_HF} in the quantum and classical regimes, for different temperatures and interactions, see also Appendix \ref{app_quantum_TBA} for the quantum formula.
In the classical model and at thermal equilibrium, the expectation value of the cosine of the phase is a function only of the adimensional coupling $\beta Mc^2$, while this scaling is absent in the quantum case: the classical curve is contrasted with the quantum result for decreasing value of $g$. 
In the high temperature limit 
\be
\beta M c^2\ll  1\, ,
\ee
the semiclassical limit is attained regardless the value of $g$, hence without the need of fulfilling Eq. \eqref{eq_semiclassic}. Indeed, in this case sine-Gordon flows to the massless point and the phase field $\phi$ has arbitrary values $\langle\cos\phi\rangle=0$. In the absence of a mass scale, the validity of semiclassics is determined solely by the lengthscale $\ell_\text{UV}$.
As one decreases the temperature, the value of $g$ becomes more relevant and quantum effects can be seen.

In the low-temperature regime $\beta Mc^2\gg 1$, the phase is locked to one of the minima and $\langle 1-\cos\phi\rangle$ decays: in the classical SG, it approaches zero with a power law $\sim 1/(\beta M c^2)$ due to Rayleigh-Jeans distribution of classical radiative modes \cite{Koch2023}, while the finite mass gap of the quantum model results in an exponential decay to a small plateau, corresponding to the ground state expectation value.

The transition from power law to exponential decay is pushed to lower and lower temperatures as one goes deeper into the semiclassical regime. Quantum curves are obtained by numerically solving the quantum TBA equations up to $\xi^{-1}=60$: going beyond this value is challenging, due to the increasing number of breathers in the spectrum (see Appendix \ref{app_quantum_TBA}). 
We also show the extrapolation to $\xi^{-1}=240$, assuming a parabolic convergence in $\xi$ to the classical limit: as we discuss in Section \ref{sec_experiment}, this value is relevant for the current experimental realizations.

We analyzed the quantum-classical correspondence on equilibrium thermal states: drastic nonequilibrium protocols may trigger genuine quantum effects and the validity of the semiclassical approximation needs to be properly assessed case-by-case.

\begin{figure}[t!]
\begin{center}
\includegraphics[width=1 \columnwidth]{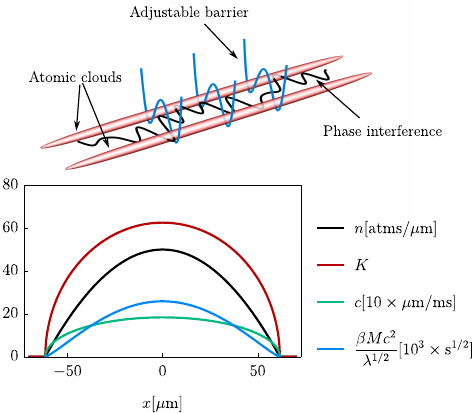}
\end{center}
\caption{\textbf{The coupled-condensates experiment.} Top: Sketch of the sine-Gordon realization with coupled quasicondensates.
Bottom: Typical profiles of quantities relevant for the emergent sine-Gordon. From top to bottom in the legend: atom density $n$, Luttinger parameter $K$, sound velocity $c$ and kink rest energy normalized to the inverse temperature and square-root of tunneling rate. In this caption, we restore the Planck constant for convenience.
These curves are obtained using the weakly-interacting approximations $K=\pi/\sqrt{\gamma}$ and $c\simeq \frac{\hbar n}{m_\text{Rb}}\sqrt{\gamma}$ (where $m_\text{Rb}$ is the mass of ${}^{87}\text{Rb}$) and with the typical parameters of Ref. \cite{Schweigler2017}: longitudinal harmonic trap with $\omega=2\pi\times 6.7 \text{Hz}$, $-2/a_{1d}=a_{3D}m_\text{Rb}\hbar\omega_{\perp}$, where $\omega_\perp=2\pi\times 1.4 \text{KHz}$ and $a_{3D}$ is the three-dimensional scattering length of Rubidium in s-wave. The density profile is computed in Thomas-Fermi approximation and we choose as typical temperature $50\text{nK}$. We set $n_\text{bulk}\simeq 50 \text{atms}/\mu \text{m}$, resulting in $\approx 3400$ atoms in each well.
}\label{Fig_3}
\end{figure}

\section{Sine-Gordon from coupled condensates}
\label{sec_experiment}
A versatile tabletop simulator of sine-Gordon is realized in Vienna \cite{Schweigler2017}, based on the blueprint proposed by Gritsev \emph{et al.} in 2007 \cite{Gritsev2007}.
In this setting, two identical quasi-onedimensional condensates  of ${}^{87}$Rb atoms are positioned close together, allowing weak tunneling of particles from one to the other.
Neglecting the tunneling in a first approximation, at low temperatures each condensate is well described by Luttinger-Liquid and fully-characterized by the sound velocity $c$ and Luttinger parameter $K$. By introducing a non-trivial tunneling rate $\lambda$ among the condensates, sine-Gordon emerges as the effective dynamics for the relative phase $\phi$ among the condensates. However, the inhomogeneous profile of the atomic cloud gives an effective spatial dependence of the sound velocity and Luttinger parameter \cite{Cazalilla2004}, resulting in an inhomogeneous sine-Gordon Hamiltonian
\be\label{eq_SGexp}
H=\int\dd x\, \frac{ c(x)}{2}\left(\frac{2\pi \Pi^2}{K(x)}+\frac{ K(x)(\partial_x\phi)^2}{2\pi}\right)-2\lambda n(x)\cos\phi\, ,
\ee
where $n(x)$ is the atom density profile. The potential $\cos\phi$ is proportional to the local atom density only in the weakly-interacting regime of the condensate, at strong interactions renormalization effects play a role \cite{shashi2011,shashi2012}. By a quick inspection with Eq. \eqref{eq_HSG}, the correspondence with the canonical sine-Gordon parameters can be made. In particular, it holds $g^2=2\pi/K$.

The sound velocity and Luttinger parameter are determined by the atomic cloud: while some degree of tunability can be achieved by acting on the particles' density and confining potentials, different parameter ranges and nonequilibrium protocols are most conveniently explored by acting upon the tunneling barrier $\lambda$, which can be tuned almost at will and modifies the mass scale of the field theory. Mass quenches are the canonical way to explore nonequilibrium in the experiment \cite{Langen2013,Kuhnert2013,Gring2012,Langen2015,Rauer2018,Schweigler2021}.

While it has been suggested that the sine-Gordon approximation may be not valid for far-from-equilibrium protocols like condensates-splitting \cite{Yuri2019,Yuri2020a,Yuri2020b}, the general consensus is that this is a reliable approximation at equilibrium \cite{Schweigler2017}. 
However, to the best of our knowledge, an exhaustive quantitative analysis of the impact of further corrections to the effective sine-Gordon dynamics, and in particular the coupling of the relative phase with symmetric degrees of freedom such as density fluctuations \cite{Yuri2020b}, is yet to be carried out as a function of the experimental parameters. 
Here, we are interested in the idealized scenario where corrections to Eq. \eqref{eq_SGexp} can be entirely neglected, and investigate the validity of the semiclassical treatment.
For this purpose, the typical numbers of the experiment are needed: in Fig. \ref{Fig_3} we show the space profile of the sine-Gordon couplings expected for a typical experimental configuration from Ref. \cite{Schweigler2017}.

Each of the two condensates is well-described by the one-dimensional interacting Bose gas Hamiltonian, also known as Lieb-Liniger (LL) model \cite{Olshanii1998,Bouchoule2022}, which is also integrable. In the LL model, particles feel a contact repulsive interaction parametrized by the one-dimensional scattering length $a_{1\text{d}}$. The adimensional quantity $\gamma=2/(|a_{1\text{d}}|n)$ is the relevant parameter to describe the different regimes of the Bose gas.
The Luttinger parameter and sound velocity depend on both the scattering length and the density of the gas and can be computed exactly \cite{Cazalilla2004}. For small $\gamma$, the approximation $K\simeq \pi/\sqrt{\gamma}$ holds, whereas in the opposite regime one has $K\to 1$. Typical parameters for the Vienna's experiment \cite{Schweigler2017} are $a_{1d}\simeq -16 \mu\text{m}$ (some tunability can be achieved acting on the transverse trap) and a bulk density $n_{\text{bulk}}\gtrsim 50 \text{atms}/\mu\text{m}$, yielding $\gamma\lesssim 2.5\times 10^{-3}$. 
Hence, in the bulk of the atom cloud, one has a Luttinger parameter $K\gtrsim 60$,  with the result that sine-Gordon hosts $\gtrsim 240$ different species of breathers and it is very likely to be realized in the semiclassical regime, depending on the temperature. 

For typical experiments one has $T\simeq 11-50\text{nK}$ \cite{Schweigler2017}, but the temperature itself is not very informative, since it has to be compared with the  tunable mass scale of the field theory. A better quantifier is $\langle 1-\cos\phi\rangle$, which is routinely probed in the experiment by matter-wave interferometry \cite{Schumm2005,Hofferberth2007,Yuri2018}. By looking at Fig. \ref{Fig_1} (bottom), we see that for $4K=8\pi/g^2\simeq 240$, a conservative estimation for the validity of the semiclassical regime is $\langle 1-\cos\phi\rangle \gtrsim 0.5$: for stronger tunneling, the extrapolation we used in Fig. \ref{Fig_1} (bottom) to reach $K\simeq 240$ is not reliable. The validity of semiclassics likely stretches beyond this point, but this needs to be checked with a more careful analysis.

As discussed in Section \ref{sec_1_a}, quantum effects may be important for correlation functions at short distances: using in the UV cutoff $\ell_\text{UV}=c\beta$ as the most conservative estimation the coldest temperature $T=10\text{nK}$ and the bulk sound velocity $c\simeq 2 \mu \text{m}/\text{ms}$, one obtains $\ell_\text{UV}\simeq 1.4 \mu\text{m} $ (upon restoring the correct dimension by inserting the Planck constant $\hbar$), which is below the current experimental resolution $\simeq 3\mu\text{m}$ \cite{schweigler2019}. Hence, quantum effects are most likely hard to see.

While deeply in the bulk the Luttinger parameter $K$ is very large and thus semiclassical physics emerges, this may be more questionable at the boundaries of the trap, where $K\to 1$ and the number of  species of breather diminishes. However, the kink mass scale is also decreasing by approaching the boundaries of the trap, thus opening to the possibility that, before that quantum effects may become important, sine-Gordon flows to the massless limit securing the validity of the semiclassical approximation. By computing the classical kink mass from Eq. \eqref{eq_SGexp}, and by using the weakly-interacting approximation for the sound velocity $c\propto 1/\sqrt{n}$, one obtains the scaling $\frac{M(x) c^2(x)}{K^{5/2}(x)}=\text{constant}$, where $M(x)$ is the soliton mass computed in position $x$ within the local density approximation. If, for example, the temperature in the bulk is set to be $0.25\%$ of the soliton rest energy and one assumes $K_{\text{bulk}}=60$, at $K(x)=20$ (corresponding to $\xi^{-1}=79$) one has $\beta M(x)c^2(x)\simeq 0.25$ which, accordingly to Fig. \ref{Fig_1}, is both deeply in the semiclassical limit and in the high-temperature regime. 

It is interesting to contrast these conclusions with the interacting Bose gas \cite{Bouchoule2022}: in this system, at the edges of the trap quantum effects are enhanced and semiclassical approximation breaks down. In this case, a quantum treatment is crucial for nonequilibrium scenarios where quantum effects propagating from the edges may affect the bulk itself.
This is not the case in sine-Gordon since, as we discussed, one can expect the semiclassical limit to describe the whole atomic cloud.

\begin{figure}[t!]
\begin{center}
\includegraphics[width=1 \columnwidth]{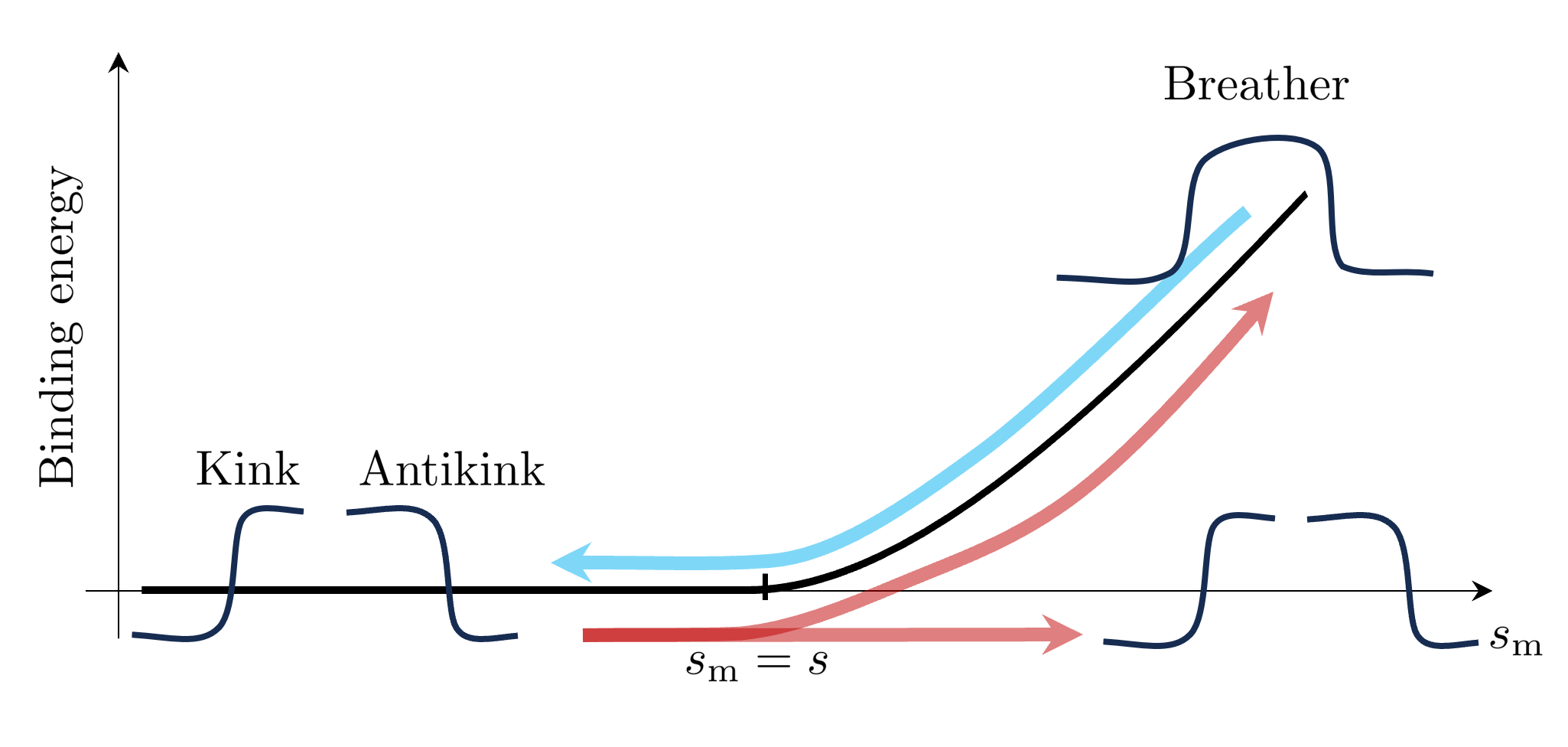}
\end{center}
\caption{\textbf{Phenomenology of breathers' melting and kink-antikink binding.} Let us consider a breather of species $s$ with binding energy $\epsilon_K(\theta)+\epsilon_{\bar{K}}(\theta)-\epsilon_s(\theta)$ and assume the interaction $\sm$ is decreasing in the reference frame of the moving breather. Due to the change of the background interactions $g_{t,x}$, the breather's binding energy changes in space and time until the special condition $s_m=s$ is met: at this point, the binding energy vanishes, and the breather melts in a pair of kink and antikink with the same rapidities (blue arrow).
The reverse process is also possible (red arrow): a pair of neighboring kink and antikink with equal rapidity can bond in a breather of species $s$ at the special point $\sm=s$. In this case, the binding is not unavoidable and the pair can also proceed as unbond particles. We stress that, since the breather's spectral parameter $s$ is continuous, for each value of $\sm$ there is always the possibility of the kink-antikink pair to form a boundstate, provided $\sm$ is increasing.
 }\label{Fig_4}
\end{figure}
\section{Generalized hydrodynamics}
\label{sec_ghd}

Motivated by the discussion on the experimental realization of the sine-Gordon model, we consider an inhomogeneous and time-dependent Hamiltonian
\be\label{eq_IHSG}
H=\int\dd x\, c_{t,x}\left\{\frac{g_{t,x}^2\Pi^2}{2}+\frac{(\partial_x\phi)^2}{2 g_{t,x}^2}+\frac{m_{t,x}^2 c_{t,x}^2}{g_{t,x}^2}(1-\cos \phi)\right\}\, ,
\ee
which we interpret classically. 
Ultimately, the most relevant case for the coupled-condensate implementation is the situation where $g_{t,x}$ and $c_{t,x}$ are constant in time, but inhomogeneous in space, while $m_{t,x}$ is also time-dependent. Nonetheless, we aim to address the general case within the framework of Generalized Hydrodynamics \cite{Doyon2016,Bertini2016,specialissueGHD}.

GHD is the hydrodynamic theory of quantum and classical models that are locally, in space and time, described by an integrable Hamiltonian. As any hydrodynamic theory, it requires a separation of scales: inhomogeneities happen on large scales, such that the system can be thought to be locally described by a steady state of the local integrable dynamics, approximated as if it was homogeneous. 

In the simplest setup when the initial state is allowed to be inhomogeneous, but the Hamiltonian is translational invariant, the dynamics can be understood as quasiparticles moving ballistically with a renormalized velocity due to the scattering shift $v(\theta)\to v^\eff(\theta)$ \cite{Doyon2016,Bertini2016}. This setting has been extensively studied in many models, including sine-Gordon in the quantum \cite{Bertini2019,Nagy2023} and classical regimes \cite{Koch2023}, but it is not enough to address the generic case of interest for us. First of all, inhomogeneities in the Hamiltonian are known to induce force terms: one should treat separately inhomogeneous potentials that couple to conserved quantities \cite{Doyon2017} (see also Ref. \cite{Durnin2021}) and those that come from more generic inhomogeneities \cite{Bastianello2019int}, which is the case for the Hamiltonian \eqref{eq_IHSG}.

Even more crucially, the spectrum of integrable models is very sensitive to interactions and, by changing the latter in space or time, the system could be locally described by a very different quasiparticle content, as it has been found for the one-dimensional Bose gas \cite{Koch2021,Koch2022} and the XXZ spin chain \cite{Bastianello2019}.
This feature pertains to the classical SG model as well: this is evident from the spectrum of the quantum theory (see Fig. \ref{Fig_1}) where breathers are quantized. At special values of the interactions $g$, the breather's energy becomes degenerate with a kink-antikink pair and binding/unbinding is possible. Even though this is a priori less clear in the continuum spectrum of the classical model, it turns out it inherits the same behavior as the quantum case.

Before presenting the GHD equations and their derivation, we discuss the physics underneath, see also Fig. \ref{Fig_4}.
Let us consider a breather of species ``$s$" traveling in the system: in quantum integrable models with multi-components, the rule of thumb is that excitations travel in the system as stable particles without changing their internal quantum number. 

In the classical case, the phase space is continuous and one can always reparametrize the internal degree of freedom $s$ in a space-time dependent manner, blurring the notion of a ``good quantum number".
In this case, quantum mechanics comes to aid and resolves this ambiguity: as we discussed in Section \ref{sec_1_a}, the quantum classical correspondence is $s\leftrightarrow n$, with $n$ the integer quantum number. Hence, the tagged excitation travels in the system by conserving its species $s$.

This has deep consequences on the stability of the breather, since it may be that the latter is eventually pushed out of the region $s\in[0,\sm]$ and it is no longer allowed in the spectrum. This signals instability: when the breather hits the boundary $s=\sm$ its binding energy vanishes, and the breather melts in a kink-antikink pair.
The opposite process is also possible: if a kink-antikink pair travels in a region of increasing $\sm$, it may bind in a breather \emph{or} continue as a pair of unbound particles.

These physical considerations are made exact by the GHD equations, valid at the Euler scale: we postpone their derivation until Section \ref{sec_derivation}, discussing first the physics and providing numerical benchmarks.
The GHD equations are better expressed in an infinitesimal form with the time-space dependence made explicit $\vartheta_\gI(\theta)\to \vartheta_\gI(t,x,\theta)$
\begin{eqnarray}\label{eq_GHDB}
\vartheta_{s}(t+\dd t,x,\theta)&=&\vartheta_{s}(t,x-\dd t v_s^\eff,\theta-\dd t F_s^\eff)\\
\nonumber
\vartheta_K(t+\dd t,x,\theta)&=&\vartheta_K(x-\dd t v^\eff_K,\theta-\dd t F_K^\eff)+\\ \nonumber
-\dd t (&\partial_t\sm&+v^\eff_{s=\sm}\partial_x\sm)2\vartheta_{s=\sm}(t+\dd t,x,\theta)\, .
\end{eqnarray}
Above, we omit the equations for the antikinks which is the same as for the kinks' case.
The GHD equations must be solved together with the boundary condition
\be\label{eq_boundary}
\vartheta_{s\ge \sm}(t,x,\theta)=\vartheta_K(t,x,\theta)\vartheta_{\bar{K}}(t,x,\theta)\, ,
\ee
which must be used to interpret shifts in Eq. \eqref{eq_GHDB} whenever the infinitesimal translation brings the filling in a region where $s\ge\sm$.
Notice that, on the left-hand side of the GHD equation for the kinks, one has the breather's filling computed at $t+\dd t$ rather than $\dd t$, as one may have naively expected. This is to ensure the correct boundary conditions: if $\dd t(\partial_t \sm +\partial_x\sm v^\eff_{s=\sm})>0$, then the equations Eq. \eqref{eq_GHDB} together with the boundary condition Eq. \eqref{eq_boundary} implies $\vartheta_{s=\sm}(t+\dd t,x,\theta)=\vartheta_K(t,x-\dd t v^\eff_K,\theta-\dd t F^\eff_K)\vartheta_{\bar{K}}(t,x-\dd t v^\eff_{\bar{K}},\theta-\dd t F^\eff_{\bar{K}})$. In this case, the GHD equation describes the process where pairs of kinks and antikinks are fusing into breathers, therefore these excitations are removed from the (anti)kink's filling, and become breathers.
In the opposite case $\dd t(\partial_t \sm +\partial_x\sm v^\eff_{s=\sm})<0$, the GHD equation describes the process where breathers become unstable and split into kinks-antikinks pairs. 

In Eq. \eqref{eq_GHDB} the effective velocity $v^\eff$ and force $F^\eff$ appear: they are meant to be computed at the same position $x$ and rapidity $\theta$ of the filling functions. The definition of the effective velocity is the usual one \cite{Doyon2016,Bertini2016} $v^\eff_\gI=(\partial_\theta\epsilon_\gI)^\dr/(\partial_\theta p_\gI)^\dr$.
The effective force is instead defined as $F^\eff_\gI=(f_\gI^\dr+\Lambda_\gI^\dr)/(\partial_\theta p_\gI)^\dr$ \cite{Bastianello2019int}, where in vectorial notation one defines
\begin{eqnarray}
f&=&-\partial_t p+\partial_t\sm   \partial_{\sm}\Theta\star[\vartheta \partial_\theta \epsilon^\dr]\, , \\
\Lambda&=&-\partial_x \epsilon+\partial_x\sm  \partial_{\sm}\Theta\star[\vartheta \partial_\theta p^\dr] \, .
\end{eqnarray}
Above, the energy and momenta have a parametric space-time dependence due to the inhomogenous couplings.
In quantum systems, the kernel $\partial_{\sm}\Theta$ is the derivative of the scattering phase with respect to the interaction. In the classical theory the same definition holds, provided one uses the expression for the classical scattering phase, which can be obtained as the limit of the quantum one \cite{Koch2023} or equivalently integrating the scattering shift $\Theta_\gI(\theta)=\int^{\theta}\dd\theta' \varphi_\gI(\theta')$. The classical scattering phase for breathers reads
\be
\Theta_{s,s'}(\theta)=\int_0^{\min(s,s')}\dd\tau G\left(\tfrac{|s-s'|+2\tau}{4\sm/\pi},\theta\right)+G\left(\tfrac{2\sm-s-s'+2\tau}{4\sm/\pi},\theta\right)\, ,
\ee
with $G(x,\theta)=4 \arctan\left(\tanh(\theta/2)/\tan x\right)$. Similarly to the scattering shift $\varphi$, the other components can be obtained through the proper limits $\Theta_s(\theta)=\frac{1}{2}\lim_{s'\to \sm}\Theta_{s,s'}(\theta)$ and $\Theta(\theta)=\frac{1}{2}\lim_{s\to \sm}\Theta_{s}(\theta)$.
It should be stressed that, when computing the breather-kink and kink-kink components of $\partial_\sm \Theta$, the derivative should be taken after the limit $s\to \sm$.

As a sanity check for the hydrodynamic equations \eqref{eq_GHDB}, it is possible to show that thermal equilibrium states in a local density approximation, i.e. where the filling is obtained by solving the TBA \eqref{eq_TBA} with a constant temperature, but plugging-in the inhomogeneous couplings, are indeed stationary states for Eq. \eqref{eq_GHDB} as it should be. The check passes through lengthy but straightforward manipulations of the GHD and TBA integral equations and we omit it.

\subsection{Boundary conditions for the coupled-condensates realization}
\label{sec_BC}

So far, we focused on the bulk GHD equations, but the proper boundary conditions should be imposed for finite-systems, such as the quasicondensate implementation. Boundaries generally break integrability, which is in turn preserved only by very special conditions \cite{Ghoshal1994}. We now discuss as open boundary conditions
\begin{eqnarray}\label{eq_BC}
\vartheta_s(t,x=0,\theta)&=&\vartheta_s(t,x=0,-\theta)\, ,\\
\nonumber\vartheta_{K}(t,x=0,\theta)&=&\vartheta_{\bar{K}}(t,x=0,-\theta)\, ,
\end{eqnarray}
 are the correct choice for the quasicondensate experiment. Above, $x=0$ is meant to be the left boundary of the atomic cloud and similar boundary conditions must be imposed on the other edge. Notice kinks are reflected into antikinks upon scattering with the boundaries.

One can obtain the conditions \eqref{eq_BC} with the following reasoning: at the edges of the atomic cloud, the sine-Gordon mass scale vanishes (see Fig. \ref{Fig_3}) and one can approximate the dynamics with the massless scalar boson, in the inhomogeneous background due to the velocity field $c(x)$ and the Luttinger parameter $K(x)$. We make the further approximation that at the boundaries $K(x)=1 $, thus obtaining the Hamiltonian describing an inhomogeneous conformal field theory with open boundary conditions for the phase field \cite{Dubail2017}: in this dynamics, left and right movers are decoupled and incoming wavepackets scattering with the boundary are fully reflected back with no changes in their shape, hence the boundary conditions \eqref{eq_BC}.
\begin{figure*}[t!]
\begin{center}
\includegraphics[width=1 \textwidth]{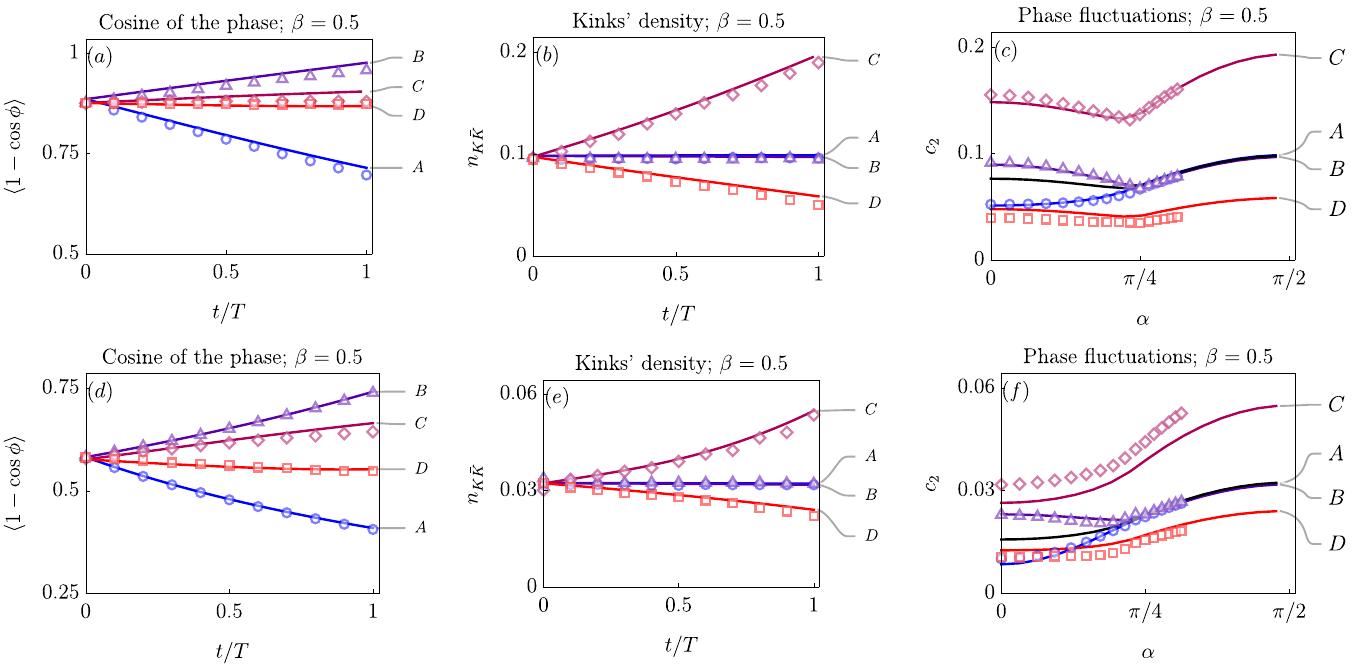}
\end{center}
\caption{\textbf{Benchmark of Generalized Hydrodynamics with Monte-Carlo.} Monte Carlo data (symbols) are compared with GHD predictions (solid lines) for different nonequilibrium protocols. Monte Carlo data are obtained by discretizing the field on a finite grid of $N$ points and lattice spacing $a$: here we only show the case $a=0.125$, $N=2^{13}$ and $T=1000$ for all cases, except protocol $C$ for $\beta=0.5$. In this case, we observe a worse convergence with the lattice spacing and show data for $N=2^{13}$ and $a=0.0625$, yet a small discrepancy from GHD is evident in panel $(f)$. See Appendix \ref{app_montecarlo} for a convergence analysis. For each dataset, we run the Monte Carlo on $20$ independent cores collecting approximately $200$ samples in each one, we use the mean as the most representative value, and we estimate the error with the mean square displacements. Errorbars within one sigma are negligible on the plot scale, and are omitted. GHD data are collected for different discretization grids: overall, we experience that convergence is challenging. We consider three different discretizations and use a linear extrapolation in the inverse of the number of points of the grid. See Appendix \ref{app_GHDdiscretization} and Appendix \ref{app_montecarlo} for further details.
We consider initial conditions with two different temperatures (top row $\beta=0.25$, bottom row $\beta=0.5$) and initial parameters $c=g=m=1$. We consider four different protocols ($c=1$ in all of them):
Protocol A: $m(t)=1+t/T$, $g=\text{const.}$. Protocol B: $m(t)=1-0.5 t/T$, $g=\text{const.}$. Protocol C:  $g^2(t)=1+t/T$, $m(t)=g^2(t)$. Protocol D:  $g^2(t)=1-0.5t/T$, $m(t)=g^2(t)$.
From left to right we show: the evolution of the the cosine of the phase \eqref{eq_HF}, the density of topological excitations $n_{K,\bar{K}}=c_2(\pi/2)$ \eqref{eq_phase_c} and finally the full kink spectroscopy at the end of the protocol $t=T$, obtained by evaluating $c_2(\alpha)$ for different angles \eqref{eq_phase_c}. In the plot for $c_2(\alpha)$, we give as a reference the same quantity computed on the initial condition (solid black line).
 }\label{Fig_5}
\end{figure*}

It is worth emphasizing that in this argument is crucial considering that the Luttinger parameter approaches a finite values at the boundaries: one could have naively extended the validity of the weakly interacting approximation $K\simeq \pi/\sqrt{\gamma}$ to the whole trap, finding in this case a Luttinger parameter that vanishes by approaching the boundaries. This naive approximation leads to incorrect boundary conditions, not describable with Eq. \eqref{eq_BC}, see Appendix \ref{app_boundary}.

\subsection{Numerical benchmarks}
\label{sec_numerics}

To benchmark the validity of the GHD equations, we consider large-scale Monte Carlo simulations of the classical sine-Gordon model. Initially thermally-distributed field configurations are sampled with the Metropolis-Hasting method \cite{Metropolis1953,Hasting1970}, then each field configuration is deterministically evolved with the equation of motion, and finally observables are obtained by averaging over the initial conditions. A short summary of the method is provided in Appendix \ref{app_montecarlo}.
Our main goal is to test the validity of GHD in the presence of force terms and particles' recombination, therefore we focus on cases where the Hamiltonian is explicitly space-time dependent: simpler transport settings with Homogeneous Hamiltonian have been already crosschecked in Ref. \cite{Koch2023}.

To solve the GHD equations \eqref{eq_GHDB}, we resort to numerical integration as well: however, we experienced numerical implementations of the GHD equations to be rather challenging. This is due to the large phase space $(\theta,s)$, to the singularities present in the kernels $\varphi$ and $\partial_\sm\Theta$, and to the singular behavior of the filling functions and root density at small $s$. In Appendix \ref{app_GHDdiscretization}, we discuss our discretization method and the aforementioned difficulties.

Even though both the GHD equations and the ab-initio Monte-Carlo simulations can address fully inhomogeneous and time-dependent protocols, for the sake of improving numerical stability and pushing the discretization, we focus on homogenoeus time-dependent couplings. In particular we consider changes in the mass $m$ and interaction $g$ (the case of a changing light velocity $c$ is analog to a mass change, plus an overall energy rescaling).

The energy is not a good observable to be looked at, due to its well-known UV divergence (see eg. Ref. \cite{Bastianello2018}), hence we focus on other quantities whose TBA expression is known: these are the already-mentioned $\langle 1-\cos\phi\rangle$ and the asymptotic two-point  connected correlator of the phase
\be\label{eq_phase_c}
c_2(\alpha)=\lim_{\ell\to\infty}\frac{\langle( \phi(t=\ell c^{-1}\cos\alpha,x=\ell\sin\alpha)-\phi(0,0))^2\rangle_\text{c}}{\ell (2\pi)^2}
\ee
where $\langle...\rangle_\text{c}$ stands for the connected part of the correlator. An analytic result for this observable has been recently obtained in Ref. \cite{Delvecchio2023}: in the classical sine-Gordon and in those states with equal root densities of kinks and antikinks $\rho_K(\theta)=\rho_{\bar{K}}(\theta)$, the simple result holds
\be
c_2(\alpha)=2\int \dd\theta \rho_K(\theta)|c^{-1}v^\eff_K(\theta)\cos\alpha-\sin\alpha|\, .
\ee
The correlator $c_2(\alpha)$ is very useful since it allows for direct spectroscopy of the population of kinks as a function of the effective velocity: in particular, we can detect changes in the total density of kinks $n_{K\bar{K}}=c_2(\pi/2)$ due to interaction changes, as predicted by the GHD equations \eqref{eq_GHDB}.

In Fig. \ref{Fig_5}, we focus on nonequilibrium protocols where, starting with thermal states, the interaction and mass scale are slowly changed on a time scale $T$: without loss of generality, we take the initial parameters $c=m=g=1$, different initial temperatures are considered. For the sake of concreteness and to have an appreciable population of kinks, we focus on high ($\beta=0.25$) and intermediate ($\beta=0.5$) temperatures.

For each initial condition, we consider four different protocols A,B,C and D: in two of these, we change the bare mass on a linear ramp $m(t)=1+(m_f-1)t/T$ while keeping the interaction $g$ fixed, where $T$ tunes the time-scale of the protocol and the final value of the mass is increased ($m_f=2$, protocol A) or decreased ($m_f=0.5$, protocol B). In these protocols, we expect $\langle\cos\phi\rangle$ to change. Kinks can also be accelerated due to force terms giving a non-trivial evolution of phase correlations \eqref{eq_phase_c}, however the total kink density $n_{K,\bar{K}}$ remains constant.

This is indeed observed in numerical data, which compare well with the GHD predictions: in particular, we observe that by increasing the mass scale, the phase locking $\langle 1-\cos\phi\rangle$ decreases, while kinks are slowed down. The opposite trend is observed by decreasing the mass.
In the other two protocols, we instead change the interactions by keeping the soliton mass $M$ fixed: we choose to change the interactions in such a way $1/\sm$ changes linearly with time, hence $g^2(t)=1+t/T(g_f^2-1)$. Of course, for slow protocols, the result does not depend on how $g^2(t)$ is varied.

We consider the explicit cases where $g_f^2=2$ (protocol C) and $g_f^2=0.5$ (protocol D): GHD predicts that the total number of kinks changes due to the formation, or melting, of breathers. In particular, by increasing $g$ the number of kinks grows, while the opposite trend is obseved in the other protocol. Also in this case, Monte Carlo data are in good agreement with GHD, proving the correctness of the equations \eqref{eq_GHDB}.

\subsection{Derivation of the GHD equations}
\label{sec_derivation}

The effective velocity and force terms in the GHD equations \eqref{eq_GHDB} have the canonical form of all integrable models \cite{Bastianello2019int} and their validity is nowaday well-established. Some ambiguity may be present in the definition of $\partial_\sm \Theta$: indeed, we could in principle change the definition of the spectral parameter $s$ with any function of the interaction $\sm$, but deriving the scattering phase $\Theta$ in $\sm$ before or after the rescaling would lead to different results. In this case, quantum mechanics comes to an aid: we can see the classical sine-Gordon as the proper limit of the quantum model \cite{Koch2023}, where there is no ambiguity due to the quantization of the breather's spectrum, confirming the validity of our choice.
The non-trivial effect to be incorporated is the melting of breathers into kinks and the opposite process: we follow the reasoning of earlier works with similar phenomenology on bound state recombination \cite{Bastianello2019,Koch2021}. As the energy suggests, a breather becomes degenerate with a kink-antikink pair $\epsilon_{s=\sm}(\theta)=\epsilon_K(\theta)+\epsilon_{\bar{K}}(\theta)$ and it holds the same for all conserved quantities (see Ref. \cite{Koch2021} for a similar discussion), hence excitations can be moved from breathers with rapidity $\theta$ and $s=\sm$, to kink-antikink pair with equal rapidity $\theta$ without altering any conservation law.

For simplicity, we now consider the case where the state is homogeneous and $\sm$ is varied in time. The proof will be easily extended to the generic case.
We work with the equivalent formulation of GHD in the root density space: in the homogeneous case, one writes the GHD equations
\be\label{eq_ghdrootK}
\partial_t\rho_K+\partial_\theta(F^\eff_K\rho_K)=(\partial_t\rho_K)_R\, ,
\ee
\be\label{eq_ghdrootB}
\partial_t\rho_s+\partial_\theta(F^\eff_s\rho_s)=0\,,\hspace{1pc}s\in[0,\sm)\, .
\ee
In the above equations, $(\partial_t\rho_K)_R$ is a ``recombination term" yet to be determined, and proper boundary conditions for $\rho_{s=\sm}$ should be imposed:
\be \label{eq_continuity}
(\partial_t\rho_K(\theta))_R=-\partial_t\sm \rho_{s=\sm}(\theta)\, .
\ee
The analogue equation for the antikink is omitted for simplicity. Notice that with this continuity equation, the conservation of the total number of particles is ensured by construction 
\be
\frac{\dd }{\dd t}\left[\int \dd\theta \left\{\rho_{K}(\theta)+\rho_{\bar{K}}(\theta)+2\int_0^{\sm}\dd s\rho_{s}(\theta)\right\}\right]=0\, .
\ee
We first consider the case where $\partial_t\sm<0$: the domain of the spectral parameter $s\in[0,\sm]$ diminishes and breathers at the boundary $s\in\sm$ become unstable and do not have other choice than melting into a kink-antikink pair. 
Notice that in this case the root density in the interval $s\in [\sm-\dd t |\partial_t\sm|,\sm]$ is known, and thus fully determines $(\partial_t\rho_K(\theta))_R$. 
We now consider the opposite scenario, namely $\partial_t\sm>0$: here the root density is initially known on the interval $s\in[0,\sm]$, while the new root density in the interval $s\in [\sm,\sm+\dd t \partial_t\sm]$ comes from the fusion of kinks-antikinks. Hence, one must have once again \eqref{eq_continuity}, but both $\rho_{s=\sm}(\theta)$ and $(\partial_t\rho_K(\theta))_R$ are unknown. This is due to the fact that a pair of kink-antikink can bind, but it can also survive as an unbounded pair: to fix the ratio of the two processes we pick the most probable possibility, obtained by maximizing the Yang-Yang entropy \cite{takahashi2005thermodynamics}.
The classical Yang-Yang entropy has been derived in Ref. \cite{Koch2023} as the semiclassical limit of the quantum one, resulting in

\begin{multline}\label{eq_YY}
\mathcal{S}=\int \dd\theta \left[\rho^t_K \eta(\vartheta_K)+\rho^t_{\bar{K}} \eta(\vartheta_{\bar{K}})+\int_{\delta_{h}}^\sm \dd s \,\rho^t_s\eta(\vartheta_s)\right]+\\
-\log h\int \dd\theta \left[\rho_K+\rho_{\bar{K}} +2\int_{\delta_{h}}^\sm \dd s \,\rho_s\right]\, ,
\end{multline}
where  $\eta(x)=x(1-\log x)$ is the entropy density of classical solitons, and $h\to 0$ is a small regulator playing the role of a Planck constant. The light-breather cutoff $\delta_h$ must be fixed such that  $\log(h \delta_h )=1$.
The GHD equations in the absence of particle recombination are known to conserve the Yang-Yang entropy \cite{Caux2019}: by taking $\partial_t\mathcal{S}$ and using the equations \eqref{eq_ghdrootK}, \eqref{eq_ghdrootB} and \eqref{eq_continuity}, through standard manipulations one obtains that the entropy rate is entirely due to the recombination term
\be
\partial_t \mathcal{S}=\int \dd\theta\, \partial_t\sm \rho_{s=\sm}\left[\log\left(\frac{\rho_{s=\sm}/\rho^t_{s=\sm}}{\vartheta_K\vartheta_{\bar{K}}}\right)-1\right]\, .
\ee
In this equation, $ \rho_{s=\sm}$ is the variable to be determined: asking for the choice that maximizes the entropy rate $\delta (\partial_t \mathcal{S})/\delta \rho_{s=\sm}=0$, one finally obtains the simple equation
\be\label{eq_12}
\vartheta_{s=\sm}(\theta)=\vartheta_K(\theta)\vartheta_{\bar{K}}(\theta)\, ,
\ee
which fixes $\rho_{s=\sm}$ and then $(\partial_t\rho_K)_R$ thanks to Eq. \eqref{eq_continuity}, leading to $\partial_t \mathcal{S}<0$. This is expected, since forming bound states diminishes the entropy of the system. After the final equations are expressed in the infinitesimal evolution form, Eq. \eqref{eq_12} is equivalent to the boundary condition \eqref{eq_boundary}.
A few further steps are needed to reach Eqs. \eqref{eq_GHDB}: first, one recasts the GHD equations from the basis of root densities to filling functions, this can be straightforwardly done with standard manipulations \cite{Bertini2016,Doyon2016}. As a useful observation, notice that when taking $\partial_t\rho^t_\gI$ the recombination term does not matter as long as Eq. \eqref{eq_continuity} is fulfilled.
The GHD equation for the kinks in terms of the filling become
\be
\partial_t\vartheta_K+F^\eff_K\partial_\theta\vartheta_K=\frac{1}{\rho^t_K}(\partial_t\rho_K)_R\, .
\ee
Now, by a direct inspection of the definition of the total root density, one observes that $\rho_K(\theta)=\frac{1}{2}\rho_{s=\sm}(\theta)$ and conveniently replace $\frac{1}{\rho^t_K}(\partial_t\rho_K)_R=-\partial_t\sm 2\vartheta_{s=\sm}$.
The final step consists in rewriting the equations in an infinitesimal form: this is convenient since in this formulation one can introduce the infinitesimal shift on the r.h.s. of the kink's equation \eqref{eq_GHDB} that automatically takes care of the two different cases $\partial_t\sm\lessgtr 0$.
This concludes the derivation of the GHD equations in the homogeneous case.
When spatial inhomogeneities are present, one can derive the GHD equations in at least two ways: the first, more complicated, method consists in generalizing Eq. \eqref{eq_YY} to spatial inhomogeneities by adding a space integration and then taking the time derivative \cite{Koch2021}.

In this way, one would find a non-trivial entropy current induced by the recombination terms: by entropy-current maximization one obtains the GHD equations \eqref{eq_GHDB}. Rather than embarking in this analysis, we use the solution of the homogeneous case and relativistic invariance. Indeed, by noticing that $\partial_\theta \epsilon_\gI=c p_\gI$ and $\partial_\theta p_\gI=c^{-1} \epsilon_\gI$, the GHD equations can be put in an explicit covariant form \cite{Doyon2016,Bastianello2019int}. For example, allowing only inhomogeneity and time dependence in $\sm$ and keeping the other parameters constant, the GHD equations for the kinks are
\be
(\mathcal{P}_K^\mu)^\dr\partial_\mu \vartheta+(\partial_{\mu}\sm)\mathcal{F}_K^\mu\partial_\theta\vartheta=\partial_\mu\sm \mathcal{R}_K^\mu\, ,
\ee
where $\partial_\mu=(\partial_t, \partial_x)$ and $\mathcal{P}^\mu=(c^{-1}\epsilon_K,  p_K)$ and the Einstein's sum convention on repeated indexes is used. Besides the covariant momentum $\mathcal{P}^\mu$, also the force terms can be recast in a covariant form $\mathcal{F}^\mu$: by completing the GHD equations with a recombination term $\mathcal{R}_K^\mu$ and asking the result to be covariant, the spatial component of  $\mathcal{R}_K$ is fixed by the temporal one, which we computed above (see eg. Ref. \cite{Bastianello2019int} for similar computations). This concludes the derivation of Eqs. \eqref{eq_GHDB}.

\section{Discussion}
\label{sec_conclusions}

In this work, we derived the hydrodynamics of the sine-Gordon field theory with inhomogeneous couplings in the semiclassical regime, showing that particle recombination must be taken into account whenever the interaction changes in time or space. 
We discussed how there is strong evidence that the sine-Gordon model currently realized on coupled-quasicondensates \cite{Schweigler2017} is well-described by the semiclassical regime, further motivating our study.

This work is a first concrete step in studying the experiment within Generalized Hydrodynamics, but there are some challenges yet to be overcome: the first practical difficulty is the numerical algorithm used to solve the GHD equations, which must be improved. The large size of the phase space due to the continuum of breather's species and the singular behavior of the kernels make convergence difficult to attain: for this reason, we limited ourselves to benchmark the predictions of hydrodynamics in the homogeneous case, while we could not obtain a satisfactory convergence in the case of spatial inhomogeneities. We hope to consider again this problem in the future: in this perspective, flea-gas algorithms \cite{Doyon2018} could be an interesting route to explore.

One important question to be addressed is whether experimental imperfections are detrimental to sine-Gordon, especially in nonequilibrium scenarios exploring time scales long enough for the inhomogeneities to play a role: the fact that a consistent hydrodynamic picture of the whole system exists, encourages to regard sine-Gordon as the proper low-energy description, at least whenever slow protocols are considered. Condensate-splitting protocols and sudden quenches have been profusely experimentally studied \cite{Rauer2018,Langen2013,Kuhnert2013,Gring2012,Langen2015}, upon abruptly changing the tunneling barrier: however, sudden protocols can be rightfully expected to excite very large energies and break the validity of sine-Gordon approximation \cite{Yuri2019,Yuri2020a,Yuri2020b}. 

In contrast, slow modulations of the tunneling barrier fall within hydrodynamics and rein-in the excitation's energy, as predicted by GHD. This should be contrasted with the Bose Gas, i.e. the current predominant experimental platform that has been compared with GHD  \cite{Schemmer2019,Malvania2020,Moller2021,Cataldini2022,Bouchoule2022}: in this case, even sudden trap quenches are describable with hydrodynamics, due to the fact that the trapping potential couples to a local conserved charge of the model, the atoms' density in this case. Hence, sudden trap changes do not excite any short-time, and thus high energy, dynamics in the Bose gas.
This is not the case for the mass scale of sine-Gordon, hence slow protocols should be used.

Even in the case of slow modulations, the importance of microscopic corrections to the effective sine-Gordon description of the coupled-condensates, primarily due to density fluctuations \cite{Yuri2020b}, is yet to be fully analyzed.
Already on equilibrium states, a comparative analysis of the spectral functions of the coupled condensates and sine-Gordon will unveil which part of the true spectrum is correctly captured by the effective field theory, giving precise boundaries on the energy scales and wavelengths. In sine-Gordon realizations on quantum spin chains, spectral functions can be probed with tensor networks and compared with the effective field theory \cite{Wybo2022}. A similar numerical analysis can be envisaged to be possible in the quasi-condensates setups with semiclassical techniques and it is left to future work.

We hope this work will serve as the basis for future developments in the GHD of the quantum sine-Gordon model: the thermodynamics and Drude weight of this quantum field theory have been recently presented \cite{Nagy2023}, and improvements in the current experiments, or realizations of new platforms \cite{Roy2021,Wybo2022,Wybo2023}, strongly advocate for developing the GHD of the quantum model.

The main difficulty in addressing the quantum sine-Gordon is due to the fact that kink and antinkink can be both transmitted or reflected upon scattering, advocating for a solution by Nested Bethe Ansatz. In contrast, in the classical limit, the scattering is always transmissive and this difficulty is absent. In the nested part of Bethe Ansatz, sine-Gordon is described by a inhomogeneous XXZ spin chain in the planar regime \cite{Nagy2023}, which notoriously has a nowhere continuous dependence on the interactions \cite{takahashi2005thermodynamics} and makes extremely challenging developing a GHD description of interaction changes.

\bigskip

\section*{Acknowledgments}
We acknowledge Rebekka Koch for collaboration on related projects and discussion at an early stage of this work.
We thank  Frederik M{\o}ller and G\'abor Tak\'acs for useful discussions, insight on the experimental apparatus, and useful comments on the manuscript. We are grateful to Giuseppe Del Vecchio Del Vecchio, M\'arton Kormos, and Benjamin Doyon for discussions and collaboration on related topics.
We acknowledge support from the Deutsche Forschungsgemeinschaft (DFG, German Research Foundation) under Germany’s Excellence Strategy–EXC–2111–390814868.

\appendix

\section{The Quantum TBA}
\label{app_quantum_TBA}
In Fig. \ref{Fig_1} we compare the classical expectation value $\langle 1-\cos\phi\rangle$ with the quantum result. Even though the focus of this work is on the classical regime, we provide here a short recap of the quantum model and the main formulas used in Fig. \ref{Fig_1}.
Due to quantum effects, the soliton mass is heavily renormalized: its value as a function of the bare mass $m$ has been computed in Ref. \cite{Zamolodchikov1995} and reads
\be
c^2M=\left(\frac{c^3 m^2}{g^2}\frac{\pi \Gamma(1/(1+\xi))}{\Gamma(\xi/(1+\xi))}\right)^{\frac{1+\xi}{2}}\frac{2\Gamma(\xi/2)}{\sqrt{\pi}\Gamma((1+\xi)/2)}\, .
\ee
where $\xi =\left(\frac{8\pi}{g^2}-1\right)^{-1}$ and $\Gamma$ is the Euler-Gamma function. 
The scattering matrix has been exactly computed in Ref. \cite{Zamolodchikov1995}. 
When a breather collides with a kink (or antikink) with relative rapidity $\theta$, they are transmitted with scattering phase
\begin{multline}
    S_n(\theta) = \frac{\sinh(\theta) + i \cos(n \pi \xi / 2)}{\sinh(\theta) - i \cos(n \pi \xi / 2)}\times\\\prod_{k=1}^{n-1}\frac{\sin^2((n-2k)\pi\xi/4 - \pi/4 + i \theta / 2)}{\sin^2((n-2k)\pi\xi/4 - \pi/4 - i \theta / 2)}\, .
\end{multline}

Instead, when two breathers with relative rapidity $\theta$ collide, they undergo transmissive scattering with scattering matrix
\begin{widetext}
\begin{multline}
    S_{n,n'}(\theta) = \frac{\sinh(\theta) + i\sin((n+n')\pi \xi/2 )}{\sinh(\theta) - i\sin((n+n')\pi \xi/2 )}\frac{\sinh(\theta) + i\sin(|n-n'|\pi \xi/2 )}{\sinh(\theta) - i\sin(|n-n'|\pi \xi/2 )}\times
 \\
\prod_{k=1}^{\min(n,n')-1}\frac{\sin^2((|n-n'|+2k)\pi \xi/4 - i\theta /2)\cos^2((n+n'-2k)\pi \xi/4 + i\theta /2)}{\sin^2((|n-n'|+2k)\pi \xi/4 + i\theta /2)\cos^2((n+n'-2k)\pi \xi/4 - i\theta /2)}\, .
\end{multline}
\end{widetext}

Finally, we discuss the scattering among kinks and antikinks, which is more complicated. While two kinks (or two antikinks) always undergo transmissive scattering with amplitude
\be\label{eq_S0}
S_0(\theta)=-\exp\left[-i\int_0^\infty \frac{\dd t}{t} \frac{\sinh(\pi t(1-\xi)/2)}{\sinh(\pi\xi t/2)\cosh(\pi t/2)}\sin(\theta t)\right]\,,
\ee
the scattering of a kink with an antikink can result either in transmission with amplitude $S_T(\theta)=\frac{\sinh (\xi^{-1}\theta)}{\sinh((i\pi-\theta)\xi^{-1})}S_0(\theta)$ or in reflection, with amplitude $S_R(\theta)=i\frac{\sin(\pi\xi^{-1})}{\sinh((i\pi-\theta)\xi^{-1})}S_0(\theta)$. 

A general solution of the thermodynamics needs to account for both scattering channels through nested Bethe Ansatz \cite{Nagy2023}. A remarkable exception takes place at the so-called reflectionless points, characterized by interaction $\xi^{-1}\in \mathbb{N}$: if this condition is fulfilled, $S_R(\theta)=0$ and thermodynamics can be computed by standard diagonal Thermodynamic Bethe Ansatz, upon defining the quantum scattering shift as $\varphi_{q;\gI}(\theta)=-i\partial_\theta\log(S_\gI(\theta))$. For the sake of simplicity, in Fig. \ref{Fig_1} we focus on the reflectionless points: we do not report the lengthy textbook formulas of the TBA and dressing equations, the reader can refer to the literature \cite{takahashi2005thermodynamics}.

The expectation value of $\langle 1-\cos\phi\rangle$ can be computed by means of the Hellmann-Feynman theorem \cite{Kheruntsyan2003}. Let us consider a multiparticle state $|\{\theta_j\}_{j+1}^N\rangle$ on a finite volume $L$, which will be taken as a representative state of the GGE once the thermodynamic limit has been taken. There, the index $j$ runs over different rapidities and particle species. The expectation value of the Hamiltonian \eqref{eq_HSG} on this state is $\langle\{\theta_j\}_{j+1}^N|H|\{\theta_j\}_{j+1}^N\rangle=L E_0+\sum_{j+1}^N m_j c^2 \cosh\theta_j$, where $E_0$ is the ground state energy density $E_0=-\tfrac{1}{4}M_{q}^2\tan(\pi\xi/2)$ \cite{Zamolodchikov1995,Lukyanov1997}. The Hellmann-Feynmann theorem states that derivatives of the energy with respect to external parameters commute with expectation values on eigenstates, thus $\langle\{\theta_j\}_{j+1}^N|\partial_m H|\{\theta_j\}_{j+1}^N\rangle=\partial_m(\langle\{\theta_j\}_{j+1}^N|H|\{\theta_j\}_{j+1}^N\rangle)$. By noticing that $\partial_m H=\int \dd x \frac{2m c^3}{g^2}(1-\cos\phi)$ one can extract the sought expectation value. Different normal-ordering prescriptions lead to a change in the normalization of the cosine term: here, we follow the conformal field theory normalization \cite{Lukyanov1997}. Some technical, but standard manipulations are needed to account for the quantization of the rapidities in finite volume and we leave them to the literature (see eg. Ref. \cite{Bastianello2019int}).
The final result is
\begin{multline}\label{eq_QTBA}
-\frac{2m^2 c^3}{g^2(1+\xi)}\langle \cos\phi\rangle=-\frac{M_q^2}{2}\tan(\pi\xi/2)+\\
\sum_{n=1}^{\xi^{-1}-1}\int\frac{\dd\theta}{2\pi}\vartheta_n[ (c^{-1}\epsilon_n\epsilon_n^\dr-cp_n^\dr p_n]+\\
\int\frac{\dd\theta}{2\pi}\Big\{\vartheta_K[ (c^{-1}\epsilon_K\epsilon_K^\dr-cp_K^\dr p_K]+\vartheta_{\bar{K}}[ (c^{-1}\epsilon_{\bar{K}}\epsilon_{\bar{K}}^\dr-cp_{\bar{K}}^\dr p_{\bar{K}}]\Big\}\, .
\end{multline}
Notice that, in contrast with the classical regime, in the quantum case the cosine operator has a non-trivial ground state expectation value, but the above expression converges to the classical result \eqref{eq_HF} in the proper limit.
Plots in Figure \ref{Fig_1} are obtained by discretizing the quantum TBA equations and using \eqref{eq_QTBA}: we discretize the rapidity space in a uniform grid of $50$ points imposing a cutoff $|\theta|\le 4$ and reaching up to $\xi^{-1}=60$, resulting in $59$ breathers in the spectrum. For larger values of $\xi$, i.e. $\xi^{-1}\simeq 30$, we checked convergence is attained upon changing the rapidity discretization.

\begin{figure}[t!]
\begin{center}
\includegraphics[width=0.99 \columnwidth]{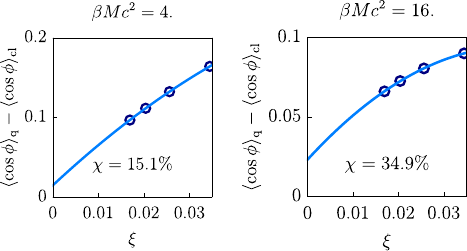}
\end{center}
\caption{\textbf{Extrapolation of $\langle 1-\cos\phi\rangle$.} In Fig. \ref{Fig_1}, we extrapolated the quantum expectation values (symbols) of the phase locking to $\xi^{-1}=240$, by assuming a parabolic convergence (solid line) to the classical result in $\xi$. The classical value is excluded from the extrapolation. In this Figure, we show two instances where the extrapolation is considered reliable (left) for sufficiently large temperatures, and the case where at low-temperature quantum data are still too far from the classical result to trust a simple parabolic extrapolation (right).}\label{Fig_ph_extr}
\end{figure}

In Fig. \ref{Fig_1}, we also show the extrapolated result up to $\xi^{-1}=240$: the convergence to the classical limit is very slow and the extrapolation cannot be trusted for all temperatures. Indeed, for the same value of $\xi$, smaller temperatures are further away from the semiclassical curve, hence we consider this compromise. We assume a parabolic convergence in $\xi$ and for each value of $\beta M c^2$ we extrapolate the expectation value on the data in the window $30<\xi^{-1}<60$. We then compare the distance of the extrapolated result from the semiclassical limit with the distance of the largest value, defining 
\be
\chi=\frac{|\langle \cos\phi\rangle_{\text{extr. to classical}}-\langle \cos\phi\rangle_{\text{classical}}|}{|\langle \cos\phi\rangle_{\xi^{-1}=60}-\langle \cos\phi\rangle_{\text{classical}}|}\, .
\ee
We decided to trust the extrapolation if $\chi<15\%$, which is realized in the window $\beta M c^2<4$: we show an explicit example in Fig. \ref{Fig_ph_extr}. For these temperatures, we add the classical limit to the data set and compute again the parabolic extrapolation until $\xi^{-1}=240$.

\section{Incorrect boundary conditions from the weakly-interacting approximation}
\label{app_boundary}

In this Appendix, we discuss the boundary conditions one would get by naively extending the weak-coupling approximation $K\simeq \pi/\sqrt{\gamma}$ to the boundaries, resulting in a vanishing Luttinger parameter instead than $K\to 1$.
For the sake of concreteness, we also make use of the weak-coupling approximation for the velocity field $c(x)= \frac{ n(x)}{m_\text{Rb}}\sqrt{\frac{2}{n(x)|a_{1d}|}}$ (in $\hbar=1$ units and $m_\text{Rb}$ is the mass of ${}^{87}\text{Rb}$) \cite{Cazalilla2004}, even though this assumption is not crucial in what follows.
By plugging these approximations in Eq. \eqref{eq_SGexp} and by neglecting the mass term, one obtains \cite{Cazalilla2004}
\be
H^\text{SG}=\int_0^\infty \dd x\,  \frac{4}{|a_{1\text{d}}|}\Pi^2(x)+\frac{1}{2}(\partial_x\phi)^2 n(x)\, .
\ee
Above, we assume the system lives on the semiaxis $x\in [0,\infty]$ and focus on the role of the boundary. At $x=0$, one must impose that the total flux of particles leaving the system is zero: the particle current is proportional to the phase gradient with a microscopic-dependent prefactor \cite{shashi2011,shashi2012} which in the weakly-interacting regime is nothing else than the atom density. Since we are already using this approximation for the Luttinger parameter, for consistency we consider the particle current in the same regime, hence the boundary condition on the microscopic field is $j(x)|_{x=0}=n(x)\partial_x \phi(x)|_{x=0}=0$.

We now study the consequences of this condition for wavepackets scattering on the boundaries.
The atom density vanishes linearly at the boundaries, therefore we approximate $n(x)=A x$ with some unimportant coefficient $A$ and write the phase field in normal modes
\be
\phi(t,x)=\int \dd \omega \, \alpha(x,\omega) b_\omega e^{i\omega t} + \alpha^*(x,\omega) b^\dagger_\omega e^{-i\omega t}
\ee
where $b_\omega$ and $b_\omega^\dagger$ are classical conjugate fields and the eigenfunctions $\alpha(x,\omega)$ satisfy
\be
\frac{\omega^2 |a_{1\text{d}}|}{8A}\alpha=-\partial_x(x\partial_x \alpha)\, .
\ee
This differential equation can be solved in terms of Bessel functions: imposing the proper boundary conditions, one obtains
\be
\alpha(x,\omega)=a(\omega) J_0\left(\omega\sqrt{\frac{x |a_{1\text{d}}|}{2A}}\right)\, ,
\ee
where $a(\omega)$ is a normalization that can be determined, but it is not important for us.
We now create a gaussian wavepacket colliding with the boundaries by asking
\be
b(\omega)=B e^{-\frac{1}{2\sigma^2}(\omega-\omega_0)^2}e^{-i\omega x_0}\, ,
\ee
where $x_0$ is the center of the wavepacket at $t=0$, the mean energy is $\omega_0$ with variance $\sigma$. $B$ is an amplitude that can be complex.

To study wavepackets, it is convenient to change coordinate as $y=\sqrt{\frac{|a_{1\text{d}}}{2A x}}$ and look at $\sqrt{y}\phi\left(x=y^2 2A/{|a_{1\text{d}}|},t\right)$. 

By using the asymptotic value of the Bessel function $J_0(x)_{x\gg 1}\simeq \sqrt{\frac{2}{\pi x}}\cos(x-\pi/4)$ and within a saddle-point approximation, one gets

\begin{widetext}
\begin{eqnarray}\label{eq_falsebc}
\sqrt{y}\phi\left(y^2\frac{2A}{|a_{1\text{d}}|},t\right)\Big|_\text{early $t$}&\simeq&\frac{ a(\omega_0)\sigma}{\sqrt{ \omega_0 }} e^{-\frac{\sigma^2}{2}(t+y-x_0)^2} \left(B e^{i\omega (t+y-x_0)-i\pi/4}+B^* e^{-i\omega (t+y-x_0)+i\pi/4}\right)\\
\nonumber\sqrt{y}\phi\left(y^2\frac{2A}{|a_{1\text{d}}|},t\right)\Big|_\text{late $t$}&\simeq&\frac{ a(\omega_0)\sigma}{\sqrt{ \omega_0 }} e^{-\frac{\sigma^2}{2}(t-y+x_0)^2} \left(B e^{i\omega (t-y+x_0)+i\pi/4}+B^* e^{-i\omega (t-y+x_0)-i\pi/4}\right)
\end{eqnarray}
\end{widetext}
The outgoing wavepacket cannot be seen as the reflection of the incoming wavepacket due to the extra phases $e^{\pm i \pi/4}$ coming from the Bessel functions, hence in this approximation \emph{one does not get reflective boundary conditions}. By propagating the incoming wavepacket in the bulk, when the cosine interaction of sine-Gordon becomes again effective, the boundary conditions \eqref{eq_falsebc} are most likely troublesome for preserving integrability. However, as we discuss in Section \ref{sec_BC}, if at the boundaries one uses $K\to 1$ rather than forcefully imposing the weakly-interacting approximation $K\simeq \pi/\sqrt{\gamma}$, fully-reflective boundary conditions are found.

\section{Numerical scheme for classical TBA and GHD}
\label{app_GHDdiscretization}

A convenient way to discretize the phase-space and compute the thermodynamics of the classical sine-Gordon has been devised in Ref. \cite{Koch2023}: we extend this scheme to hydrodynamics, which we first conveniently reparametrize.
The domain of the continuous spectral parameter labeling the breathers $s\in [0,\sm]$ is affected by interaction's changes, therefore it is convenient to define a new variable $\sigma=s/\sm\in [0,1]$. 

The filling obtained by solving the classical TBA \eqref{eq_TBA} is singular at small values of $s$, therefore it is convenient to extract the non-singular part.
To this end, we define $\bar{\vartheta}_\sigma(\theta)\equiv s^2\vartheta_s(\theta)\big|_{s=\sm\sigma}$ as the non-singular part of the filling in the breather space. For the kinks and antikinks, we define $\bar{\vartheta}_K(\theta)\equiv\vartheta_K(\theta)$ and $\bar{\vartheta}_{\bar{K}}(\theta)\equiv\vartheta_{\bar{K}}(\theta)$.
With this reparametrization, the GHD equations \eqref{eq_GHDB} and boundary condition \eqref{eq_boundary} become
\begin{figure*}[t!]
\begin{center}
\includegraphics[width=1 \textwidth]{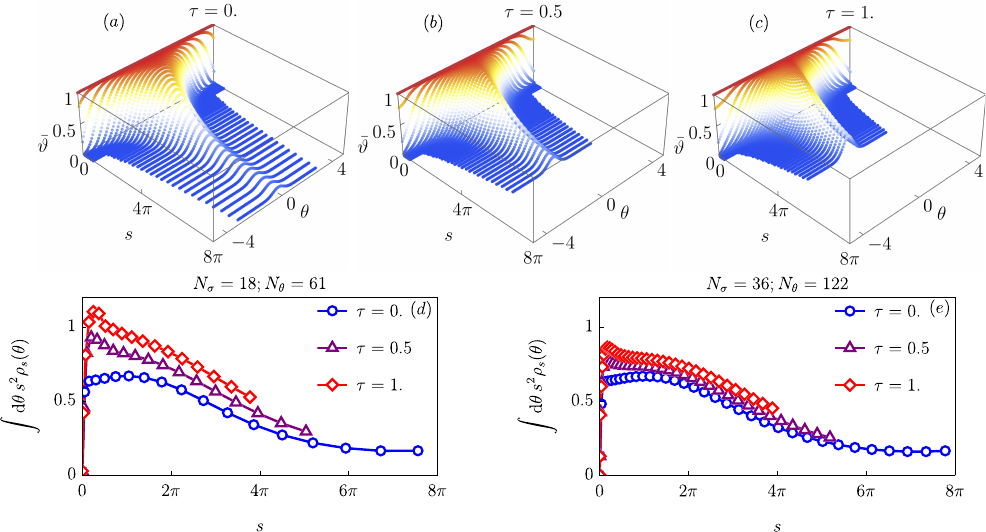}
\end{center}
\caption{\textbf{Phase space and conservation laws of GHD.} We give further details on one of the nonequilibrium protocols discussed in Fig. \ref{Fig_5}, more precisely we analyze the case where the soliton mass is kept constant and the interaction is changes as $g^2(\tau)=1+\tau$ with $\tau\in[0,1]$, see Fig. \ref{Fig_5} for further details.
Panels $(a-c)$ the nonsingular part of the breathers' filling $\bar{\vartheta}$ is shown for at different times, we plot the phase space $(\theta,s)$. As $g(\tau)$ grows, $\sm$ is reduced and breathers unbind into kinks.
Panels $(d-e)$ for two different discretizations, we plot $\int \dd\theta s^2\rho_s(\theta)$ during the evolution. For those breathers that have not been melted into kink-antikink pairs, the integrated breathers' root density is conserved by the GHD equations, but the discretization breaks the conservation law, which is improved upon increasing the number of points in the discretization.
See the main text for further discussion.}\label{Fig_app_ghd}
\end{figure*}
\begin{multline}\label{eq_app_ghd1}
\bar{\vartheta}_{\sigma}(t+\dd t,x,\theta)=\\
=\bar{\vartheta}_{\sigma+\frac{\dd t}{\sm}  (\partial_t \sm+\partial_x \sm \bar{v}^\text{eff}_\sigma)\sigma }(t,x-\dd t \bar{v}_\sigma^\eff,\theta-\dd t \bar{F}_\sigma^\eff)
\end{multline}
\begin{multline}\label{eq_app_ghd2}
\bar{\vartheta}_K(t+\dd t,x,\theta)=\bar{\vartheta}_K(x-\dd t \bar{v}^\eff_K,\theta-\dd t \bar{F}_K^\eff)+\\ 
-\dd t (\partial_t\sm+\bar{v}^\eff_{\sigma=1}\partial_x\sm)\frac{2}{ s_{\text{m}}^2}\bar{\vartheta}_{\sigma=1}(t+\dd t,x,\theta)
\end{multline}
And $
\bar{\vartheta}_{\sigma\ge 1}(t,x,\theta)=s_{\text{m}}^2\bar{\vartheta}_K(t,x,\theta)\bar{\vartheta}_{\bar{K}}(t,x,\theta)$. Above, $\bar{v}^\text{eff}$ and $\bar{F}^\text{eff}$ are nothing else than the effective velocity and force computed in the new parametrization. 
As we reparametrized the filling in terms of its non-singular part, we also define a new dressing operation \cite{Koch2023} which we denote as ``bold dressing".
This is useful because the dressing of quantities that in the original parametrization vanish linearly for $s\to 0$ (such as the energy and momentum), develops a $\sim s^2$ zero once dressed: for the sake of numerical stability, it is rather convenient to extract this prefactor. The new dressing operation accomplishes this task.

For any triplet of test functions $\{\tau_K(\theta),\tau_{\bar{K}}(\theta),\tau_s(\theta)\}$ and their reparametrization $\bar{\tau}_K(\theta)=\tau_K(\theta)$, $\bar{\tau}_{\bar{K}}(\theta)=\tau_{\bar{K}}(\theta)$ and $\bar{\tau}_\sigma(\theta)=\tau_{\sm\sigma}(\theta)$, one defines

\begin{widetext}
\be
\sigma^2\bar{\tau}^{\ddr}_{ \sigma}(\theta)=\bar{\tau}_{\sigma}(\theta)-\int \frac{\dd\theta'}{2\pi}\bar{\varphi}_{ \sigma}(\theta-\theta')[\bar{\vartheta}_K(\theta')\bar{\tau}^{\ddr}_K(\theta')+\bar{\vartheta}_{\bar{K}}(\theta')\bar{\tau}^{\ddr}_{\bar{K}}(\theta')]
-\int_{0}^1 \frac{\dd\sigma'}{\sm} \int \frac{d\theta'}{2\pi} \bar{\varphi}_{\sigma,\sigma'}(\theta-\theta')\bar{\vartheta}_{\sigma'}(\theta')\bar{\tau}^{\ddr}_{\sigma'}(\theta')\, ,
\ee
\be
\bar{\tau}^{\ddr}_K(\theta)=\bar{\tau}_K(\theta)- \int \frac{\dd\theta'}{2\pi} \bar{\varphi}(\theta-\theta')[\bar{\vartheta}_K(\theta')\bar{\tau}^{\ddr}_K(\theta')+\bar{\vartheta}_{\bar{K}}(\theta')\bar{\tau}^{\ddr}_{\bar{K}}(\theta')]
-\int_{0}^1 \frac{\dd\sigma}{\sm} \int \frac{\dd\theta'}{2\pi} \bar{\varphi}_{\sigma}(\theta-\theta')\bar{\vartheta}(\theta',\sigma)\bar{\tau}^{\ddr}_{\sigma}(\theta')\, ,
\ee
\end{widetext}
where $\bar{\varphi}_\gI$ is nothing else than the scattering kernel \eqref{eq_varphi} upon reparametrization $\sigma=s/\sm$. The connection between the bold dressing and the standard definition is $\tau_s^\dr|_{s=\sm \sigma}=\sigma^2 \tau_\sigma^\ddr$ and  $\tau_K^\dr= \tau_K^\ddr$ \cite{Koch2023}: this is rather convenient, since it means $(\partial_\theta p_s)^\dr\vartheta_s(\theta)\Big|_{s=\sm \sigma}=s^2_{\text{m}}(\partial_\theta \bar{p}_\sigma)^\ddr\bar{\vartheta}_\sigma$ and the ratio of dressed quantities can be computed equivalently with the canonical dressing and bold dressing, for example $\bar{v}_\gI^\eff=(\partial_\theta \bar{\epsilon}_\gI)^\ddr/(\partial_\theta \bar{p}_\gI)^\ddr$ and the same holds for the force terms.

We are now ready to discretize the GHD equations in the new form.
We impose a maximum cutoff in the rapidity space $\theta\in[-\theta_\text{cutoff},\theta_{\text{cutoff}}]$ and uniformly discretize this interval $\{\theta_i\}_{i=1}^{N_\theta}$. 
The $\sigma$ variable is also discretized $\{\sigma_i\}_{i=1}^{N_\sigma}$, but we keep open the possibility of considering inhomogeneous discretizations to better capture fast-changing profiles at small $\sigma$, see eg Fig. \ref{Fig_app_ghd}.
Integral equations are converted in vector equations on the discretized phase space, e.g.
\begin{multline}
\int\dd\theta' \int \dd\sigma' \bar{\varphi}_{\sigma_a,\sigma'}(\theta_i-\theta') F_{\sigma'}(\theta')\to\\
 \sum_{a=1}^{N_\sigma}\sum_{j=1}^{N_\theta} \bar{\varphi}^{\text{dis}}_{\{\sigma_a,\theta_i\},\{\sigma_b,\theta_j\}} F_{\sigma_j}(\theta_j)\, ,
\end{multline}
where $F_{\sigma'}(\theta')$ is an arbitrary smooth function, and 
\be\label{eq_varphi_dis}
\bar{\varphi}^{\text{dis}}_{\{\sigma_a,\theta_i\},\{\sigma_b,\theta_j\}}\equiv\int_{\tfrac{\theta_{j-1}+\theta_{j}}{2}}^{\tfrac{\theta_{j+1}+\theta_{j}}{2}}\dd\theta'\int_{\tfrac{\sigma_{b-1}+\sigma_{b}}{2}}^{\tfrac{\sigma_{b+1}+\sigma_{b}}{2}}\dd\sigma' \bar{\varphi}_{\sigma_a,\sigma'}(\theta_i-\theta')\, .
\ee
Notice that the scattering shift has \eqref{eq_varphi} logarithmic singularities: to correctly approximate the above integral, we first extract the singular terms of $\varphi_\gI$ and compute their integral exactly, the integral of the remaining non-singular part is approximated by the midpoint rule. Further details of this discretization, as well as some further tricks used in properly discretizing the TBA equations \eqref{eq_TBA} are reported in Ref. \cite{Koch2023} and thus we do not repeat this lengthy discussion here. Instead, we discuss the discretization of the force term, and in particular integrals where $\partial_\sm \Theta$ appears: by first taking the $\sm$ derivative and then changing parametrization to the $\sigma$ space $\partial_\sm \Theta_\gI\to \overline{\partial_\sm \Theta}_\gI$, one obtains
\begin{multline}\label{eq_Theta_app}
\overline{\partial_\sm\Theta}_{\sigma,\sigma'}(\theta)=-\int_0^{\min(\sigma,\sigma')}\dd\tau \Bigg\{\tfrac{|\sigma-\sigma'|+2\tau}{4/\pi}G'\left(\tfrac{|\sigma-\sigma'|+2\tau}{4/\pi},\theta\right)+\\
\tfrac{-\sigma-\sigma'+2\tau}{4/\pi}G'\left(\tfrac{2-\sigma-\sigma'+2\tau}{4/\pi},\theta\right)\Bigg\}\, ,
\end{multline}
where we defined $G'(x,\theta)=\partial_xG(x,\theta)=\frac{4\sinh\theta}{\cos(2x)-\cosh\theta}$. The kernel $\overline{\partial_\sm\Theta}_{\sigma,\sigma'}(\theta)$ is thus discretized similarly to Eq. \eqref{eq_varphi_dis}: we notice that the primitive with respect to the rapidity integration can be analytically taken as $IG'(x,\theta)\equiv\int^\theta \dd\theta' G'(x,\theta)=-4\log(\cos\theta-\cos(2x))$. We still have to perform the $\tau-$integral in Eq. \eqref{eq_Theta_app} and finally the $\sigma'-$ integral is discretized analogous to Eq. \eqref{eq_varphi_dis}. 

Since $IG'(x,\theta)$ has a logarithmic singularity, once the first integration over $\tau$ has been taken, the resulting expression is regular and we can approximate Eq. \eqref{eq_varphi_dis} by midpoint rule. To compute the $\tau$ integration, we first split $IG'(x,\theta)$ into the singular and non-singular part, by defining $IG'_S(x,\theta)=-4\log(\tfrac{\theta^2+(2x)^2}{2})$ and $IG'_{NS}(x,\theta)=IG'(x,\theta)-IG'_S(x,\theta)$.

By plugging this splitting into Eq. \eqref{eq_Theta_app}, the integrals with $IG'_S(x,\theta)$ can be performed analytically, while for the regular parts with $IG'_{NS}(x,\theta)$ we approximate the $\tau-$integration with the midpoint rule, by discretizing the interval $[0,\min(\sigma,\sigma')]$ with $N_\tau$ equispaced points. We experience this discretization of the integrals of the force terms is stable, even after dressing: in this respect, it is crucial to capture the $\sigma\to 0$ behavior with high precision. This is why for each integral $\tau\in [0,\min(\sigma,\sigma')]$ with the same number of points, rather than picking $\{\tau_i\}_i$ from the overall $\{\sigma_i\}_i$ discretization. This choice would have given a poor discretization of Eq. \eqref{eq_Theta_app} for small values of $\min(\sigma,\sigma')$.

Computing the discretized kernels $\bar{\varphi}^{\text{dis}}$ and $\overline{\partial_\sm\Theta}^\text{dis}$ is costly: by taking advantage of a flat discretization in the rapidity space, the cost of computing $\bar{\varphi}^{\text{dis}}$ scales as $N_\sigma\times N_\sigma\times (2N_\theta)$, while the cost of computing $\overline{\partial_\sm\Theta}^\text{dis}$ grows as $N_\sigma\times N_\sigma\times N_{\tau}\times (2N_\theta)$. However notice that, apart from an overall $\sm-$dependent factor in $\bar{\varphi}^{\text{dis}}$, the matrixes remain constants upon changing the interactions, mass, or sound velocity in sine-Gordon. Therefore, one can compute these matrixes once and for all without the need to update them at each step of the GHD evolution, saving a lot of computational effort.

\begin{figure*}[t!]
\begin{center}
\includegraphics[width=0.95 \textwidth]{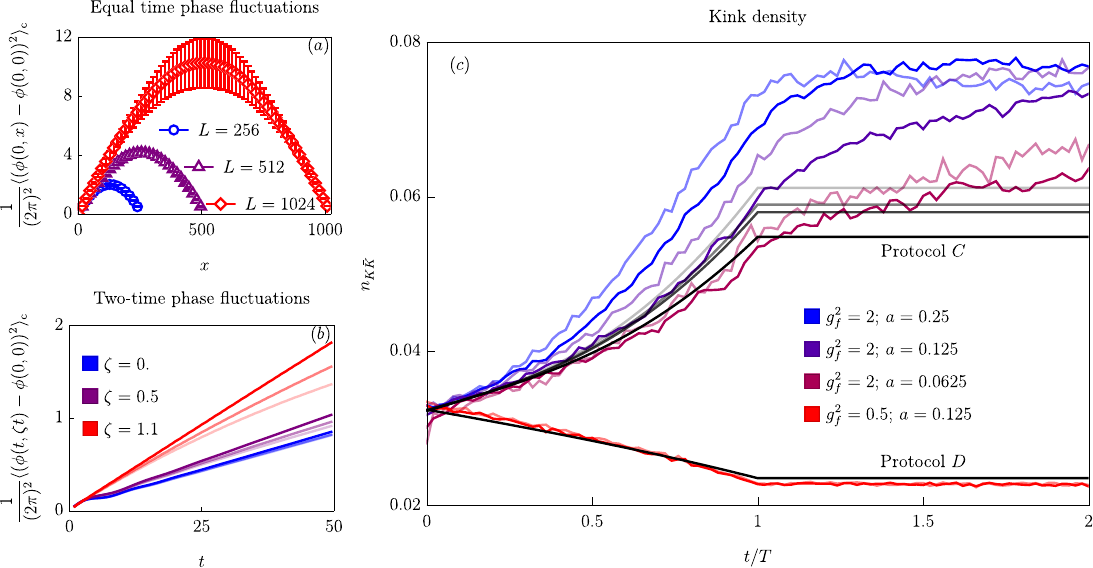}
\end{center}
\caption{\textbf{Convergence analysis of Monte Carlo data.}Panel $(a)$: we consider equal-time phase fluctuations at equilibrium on a thermal state with $m=g=c=1$ and $\beta=0.5$. We keep the lattice spacing fixed $a=0.125$ and consider three different sizes $L=a\times N$. Due to periodic boundary conditions, the phase fluctuation does not grow linearly, but bends with the approximate shape of an inverted parabola. 
Panel $(b)$: two-time phase fluctuations are shown for different choices of the ray $\zeta=x/t$, we consider the same equilibrium state, sizes and lattice spacing of panel $(a)$. Different colors refer to different rays, we use the opacity to label the different sizes (the transparency increases upon diminishing the size). Larger values of $\zeta$ are prone to slowest corrections in the system's size, but are approximately linear also at short times. Smaller values of $\zeta$ show finite-time oscillations, which makes extracting $c_2$ more challenging. Error bars are not shown, since negligible on the plot scale.
Panel (c): We consider the nonequilibrium protocol obtained by changing $g$ in the interval $t<T$ and keeping it constant afterward, while the soliton mass is kept fixed, hence considering protocols $C$ and $D$ of Fig. \ref{Fig_5} (see main text for more details). We start from thermal states $m=g=c=1$ and $\beta=0.5$, the system's size is fixed to $L=512$ and we consider three different lattice spacing (colors) and two different times $T=1000$ (dark shading) and $T=2000$ (light shading). The GHD curve is given for reference (black solid line): for Protocol C, which has the worse convergence, we show in different shading from lighter to darker the three discretizations used in GHD and their extrapolation (see Appendix \ref{app_GHDdiscretization}). In the case of Protocol D, the difference among the three discretizations is negligible: we show only the extrapolation. For $t>T$, the kink's density should remain constant: integrability-breaking effects due to finite $a$ cause instead a drift, which is reduced by diminishing $a$. Error bars are not shown for the sake of clarity.
}\label{Fig_app_montecarlo}
\end{figure*}

The discretized solution of the integral equations is then used to compute the force term and effective velocity in the GHD equations \eqref{eq_app_ghd1} \eqref{eq_app_ghd2}: we solve them by using a forward/backward first-order interpolation for translations in the phase-space, the direction of the interpolation is chosen according with the shift. To increase the stability of the time evolution, the force terms and effective velocities in Eqs. \eqref{eq_app_ghd1} \eqref{eq_app_ghd2} are computed at the midpoint $t+\dd t/2$, and the filling used in the dressing is the average between the forward and backward evolution $\frac{1}{2}(\vartheta_{t+\dd t}+\vartheta_t)$. The solution for $\vartheta_{t+\dd t}$ is then obtained recursively. 
As an example, in Fig. \ref{Fig_app_ghd} we give further details on the protocol shown in Fig. \ref{Fig_5} which we experienced having worst convergence, namely when the kink mass is kept fixed and the interaction increased $g^2(\tau)=1+\tau$ starting with the thermal ensemble with $\beta=0.5$. In the subfigures $(a-b-c)$ we show the non-singular part of the breathers' filling function $\bar{\vartheta}$ in the phase space $(\theta,s)$ for different times. The discretization shown has a cutoff in the rapidity space $\Lambda_\text{cutoff}=5$, $N_\theta=122$ and $N_\sigma=36$: the discretization in the rapidity direction is flat, while in the $\sigma$ direction we choose a parabolic discretization, which is denser at small $\sigma$. As $g$ is increased, the breather phase space $s\in [0,\sm]$ is reduced and breathers unbind into kinks-antikins pairs.
In the GHD equations, $\int \dd\theta \rho_s(\theta)$ is constant until $s<\sm$, as it is evident from Eq. \eqref{eq_ghdrootB}. This conservation law is only approximate in the discretized equations, hence it is a diagnostic tool to quantify the quality of the discretization. However, for small $s$ the root density has a divergence $\sim 1/s^2$ \cite{Koch2023}, therefore it is challenging for the discretized GHD equations to correctly capture the root density. In subfigures $(e-f)$ we show $\int \dd\theta  s^2\rho_s(\theta)$ (notice the extra prefactor $s^2$ to remove the singularity) for different times and two different discretizations. Adding more points to the discretization improves the conservation law: we considered three different discretizations (the two shown in panels $(e-f)$ and an intermediate one),  rescaling with $N_{\sigma}=\alpha\bar{N}_{\sigma}$ and $N_{\theta}=\alpha\bar{N}_{\theta}$, with $\alpha$ a scaling parameter and $\bar{N}_{\sigma}$, $\bar{N}_{\theta}$ kept fixed. In practice, we choose $\bar{N}_{\sigma}=18$ and $\bar{N}_{\theta}=61$ and $\alpha=\{1,3/2,2\}$: pushing the discretization to even smaller grids is challenging, but on the collected data we see an approximate linear scaling in $1/\alpha$, thus we extrapolate to $\alpha\to\infty$. See also Fig. \ref{Fig_app_montecarlo}. 
For protocols where the bare mass $m$ is changed, but the interaction $g$ is kept constant, convergence is much better (plots not shown).
The slow convergence of GHD is the main bottleneck to run late time and spatially-inhomogeneous hydrodynamic simulations.

\section{The Monte Carlo analysis }
\label{app_montecarlo}

In this appendix, we provide a short overview of the standard methods used to simulate the microscopic classical dynamics. 
Since we focused on time-changes of the SG coupling, we restrict ourselved to this case and thus assume homogeneity in space.
We first conveniently rescale the fields with the canonical transformation 
\be\label{eq_rescaling}
\phi(t,x)\to g(t) \phi(t,x)\hspace{1pc}\text{and} \hspace{1pc}\Pi(t,x)\to \Pi(t,x)/g(t)\,
\ee
and we discretize the system on a uniform grid with lattice spacing $a$  and periodic boundary conditions, obtaining the discrete Hamiltonian
\be\label{eq_disH}
H[\Pi,\phi]=a\sum_j \frac{c }{2}\Pi_j^2+\frac{c}{2  a^2}(\phi_{j+1}-\phi_j)^2-\frac{c^3 m^2}{g^2}\cos(g\phi_j)\, ,
\ee
where the fields obey canonical Poisson brackets $\{\phi_j,\Pi_{j'}\}=\delta_{j,j'}/a$. For the sake of simplicity, in the following we set $c=1$, while $g$ and $m$ are kept as time-dependent functions $g\to g(t)$ and $m\to m(t)$. On thermal ensembles, the distribution of $\Pi$ and $\phi$ are independent: the first follows indpendent gaussian distributions for each lattice site, while the distribution of $\phi_j$ is sampled with Metropolis-Hasting methods \cite{Metropolis1953,Hasting1970}, by suitably chosen random walks. Local updates $\phi_j\to\phi'_j= \phi_j+\delta \phi_j$ are accepted with probability $p=\exp(-\beta H[\phi'])/\exp(-\beta H[\phi])$, where $H[\phi]$ is the phase-dependent part of the discretized Hamiltonian \eqref{eq_disH}. We choose $\delta \phi_j$ as gaussianly distributed with zero mean, the variance is tuned in such a way as to achieve approximately $50\%$ acceptance ratio.

After a sufficiently long time, the Metropolis random walk converges to the thermal distribution and sampling begins: we take the field configurations generated by the Metropolis and deterministically evolve with the equation of motion obtained from the discrete Hamiltonian \eqref{eq_disH}. Observables are then computed through the evolution and averaged over the initial conditions.
The time evolution is discretized with the following method \cite{DeLuca2016}
\begin{multline}
\phi_j(t+\dd t)-\phi(t)-(\phi_j(t)-\phi(t-\dd t))=\\
\frac{\dd t^2}{ a^2}(\phi_{j+1}(t)+\phi_{j-1}(t)-2\phi_j(t))-\dd t^2\frac{m^2(t)}{g(t)}\sin(g(t)\phi_j(t))\, .
\end{multline}
The above equation is seen as an update rule for $\{\phi_j(t+\dd t)\}_{j=1}^N$, given $\{\phi_j(t)\}_{j=1}^N$ and $\{\phi_j(t-\dd t)\}_{j=1}^N$: the initial conditions are determined by the field configurations sampled from the Metropolis which fixes $\{\phi_j(t=0)\}_{j=1}^N$, and using that $\phi_j(\dd t/2)=\phi_j(0)+\tfrac{\dd t}{2}\Pi_j(0)$. This discretization method is stable if $\dd t/a<1$ and errors on energy conservation remain bounded in time and $\sim \mathcal{O}(\dd t)$, but for finite lattice size $a$, this discretization breaks integrability, in contrast with other symplectic integrators \cite{Orfanidis1978}. However, since we are interested in the continuum sine-Gordon, the limit $a\to 0$ has to be taken anyway and lattice-induced integrability terms can help in assessing the quality of the data.

Error bars are obtained by running each protocol on $20$ independent copies and considering the variance, while we take the mean as the most representative value. We take advantage of translational invariance to perform further averaging. Typically we reach good convergence by collecting $\sim 200$ samples for each copy, thus $\sim 4000$ samples: error bars show the confidence interval within one sigma, and are omitted if negligible on the scale of the plot.

In Fig. \ref{Fig_5} we showed the final processed data, now in Fig. \ref{Fig_app_montecarlo} we explicitly discuss the data analysis for some typical parameters. We experienced that the density of kinks is the quantity affected most by finite sizes and lattice spacing, hence we show the scaling of the latter: phase flucutations are computed with respect to the original phase field $\phi(t,x)$ (properly discretized) and and not on the rescaled fields \eqref{eq_rescaling}.
We first discuss finite-size corrections: in an infinite system, the phase correlation $\langle (\phi(0,x)-\phi(0,0))^2\rangle_\text{c}/(2\pi)^2$ linearly grows upon increasing the separation $x$. Instead, periodic boundary conditions force the correlator to bend to reach zero as $x=L$, with the shape of an approximated inverted parabola, as already noticed in Ref. \cite{Delvecchio2023}: in Fig. \ref{Fig_app_montecarlo} panel $(a)$, we show the equal time correlation function for different sizes. The linear growth is extracted by focusing on large distances with respect to the microscopic scale, but much smaller than the system's size. For the same reason, two-times phase fluctuations also depend on finite size corrections, but with the additional complication that short-time oscillations force us to focus on even larger separation to take a reliable linear fit: in Fig. \ref{Fig_app_montecarlo} panel $(b)$, we show the phase fluctuations as a function of time $t$, with the space separation $x=c t\tan\alpha$, for various values of $\alpha$. We keep the space (or time) windows where we focus on extracting the linear growth of the phase-fluctuations constant upon changing systems' size and perform a linear fit, then the so-obtained slope is compared for different sizes. The procedure is repeated upon changing the fit region, to ensure the fit does not depend on it: on the scale of the plots of Fig. \ref{Fig_5}, the two largest system's sizes we explored do not show appreciable differences.

Lastly, we consider the effect of a finite lattice size $a$, which is twofold. On the one hand, the UV cutoff $\sim 1/a$ must be improved as the temperature increases: this effect is mostly controlled by $\beta$ and is relatively independent on from interactions.
The second issue, is that sine-Gordon excitations (both kinks and breathers) are smooth field configurations and the lattice spacing $a$ must be much smaller than their size, which depends on interactions. 

A finite lattice spacing in Eq. \eqref{eq_disH} breaks integrability: a convenient way to quantify the role of $a$, is by looking at conservation laws of the continuum model, such as the kinks' density.
In Fig. \ref{Fig_app_montecarlo} panel $(c)$, we consider the nonequilibrium protocol shown in Fig. \ref{Fig_5} which has the worst convergence, more precisely we initialize sine-Gordon at $c=m=g=1$ for $\beta=0.5$ and change the interactions as $g^2(t)=1+(g_f^2-1)t/T $ is $t<T$ and kept constant at later times. The soliton mass is kept constant in the protocol and the final interactions are chosen as $g_f^2=2$ and $g_f=0.5^2$. These are nothing else than the protocols $C$ and $D$ respectively, shown in Fig. \ref{Fig_5}: we keep the same notation for consistency.

For $t<T$ the interactions change and the density of kinks will evolve accordingly. For $t>T$, the interactions are kept fixed and, in the continuum limit, the field evolves with a time-independent sine-Gordon Hamiltonian, thus conserving the number of kinks. Since GHD works for large $T$, one could naively expect that the agreement is improved upon increasing $T$. However, if $T$ is too large, integrability-breaking effects due to a finite $a$ are important: for increasing interactions, i.e. protocol $C$, instead of having a flat plateau we observe a drift in the number of kinks is observed for $t>T$, which is reduced for smaller $a$. For the case where interactions decrease (protocol $D$) we final plateau remains constant with good approximation, hence integrability-breaking terms are more negligible. A small difference is still visible between Monte Carlo and GHD, which is most likely due to the difficult convergence of the latter.

Notice that, at fixed $a$, there is an optimal value for $T$: if $T$ is too small, corrections with respect to GHD are expected. If $T$ is too large, integrability corrections play a role and are actually more important as $T$ is further increased.

\bibliography{biblio}

\end{document}